\documentclass[12pt,preprint]{aastex}
\usepackage{graphicx}
\usepackage{natbib}
\usepackage{lscape}

\shorttitle{T\,Tauri Jet Physics Resolved Near The Launching Region with the Hubble Space Telescope}
\shortauthors{Coffey~et~al.\,2008}

\begin{document}

\title{T\,Tauri Jet Physics Resolved Near The Launching Region with the Hubble Space Telescope
\footnote{Based on observations made with the 
NASA/ESA $\it{Hubble}$ $\it{Space}$ $\it{Telescope}$, 
obtained at the Space Telescope Science Institute, which is operated by the Association of 
Universities for Research in 
Astronomy, Inc., under NASA contract NAS5-26555.}}

\author{
Deirdre Coffey\altaffilmark{1}, Francesca Bacciotti\altaffilmark{1}, Linda Podio\altaffilmark{2}}

\altaffiltext{1}{I.N.A.F. - Osservatorio Astrofisico di Arcetri, Largo E. Fermi 5, 50125 
Firenze, Italy \email{dac, fran@arcetri.astro.it}}
\altaffiltext{2}{Dublin Institute for Advanced Studies, 31 Fitzwilliam Place, Dublin 2, 
Ireland \email{lindapod@cp.dias.ie}}

\begin{abstract}

We present an analysis of the gas physics at the base of jets from five T Tauri stars based on high angular resolution optical spectra, using the Hubble Space Telescope Imaging Spectrograph (HST/STIS). The spectra refer to a region within 100 AU of the star, i.e. where the collimation of the jet has just taken place. We form PV images of the line ratios to get a global picture of the flow excitation. We then apply a specialised diagnostic technique to find the electron density, ionisation fraction, electron temperature and total density. Our results are in the form of PV maps of the obtained quantities, in which the gas behaviour is resolved as a function of both radial velocity and distance from the jet axis. They highlight a number of interesting physical features of the jet collimation region, including regions of extremely high density, asymmetries with respect to the axis, and possible shock signatures. Finally, we estimate the jet mass and angular momentum outflow rates, both of which are fundamental parameters in constraining models of accretion/ejection structures, particularily if the parameters can be determined close to the jet footpoint. Comparing mass flow rates for cases where the latter is available in the literature (i.e. DG Tau, RW Aur and CW Tau) reveals a mass ejection-to-accretion ratio of 0.01 - 0.07. Finally, where possible (i.e. DG Tau and CW Tau), both mass and angular momentum outflow rates have been resolved into higher and lower velocity jet material. For the clearer case of DG Tau, this revealed that the more collimated higher velocity component plays a dominant role in mass and angular momentum transport. 

\end{abstract}

\keywords{ISM: jets and outflows --- stars: formation, pre-main sequence --- 
stars: individual: TH 28, DG Tau, CW Tau, HH 30, RW Aur}


\section{Introduction}
\label{introduction}

In current star formation theory, jets/outflows from a newly forming star are believed 
to transport significant amounts of energy and momentum away from the region of 
the central source (\citealp{Bally07}; \citealp{Ray07}; \citealp{Pudritz07}; \citealp{Shang07}). 
This can have a big influence on the way in which the stars form, 
because, for example, jets may drive the injection of turbulence in the parent cloud, 
thereby regulating the star formation rate. At the same time, 
they may be able to extract the excess angular 
momentum  from the accretion disk, thus allowing the matter to drift through it 
and finally accrete onto the star. Jets, therefore, are considered a fundamental ingredient 
in the star formation process. 
To fully understand the  mechanisms underlying the physics  
of these commonly observed nebulae, however, it is necessary to know the mass outflow rate, which regulates
the dynamics of the flow and, therefore, is the most important input 
parameter for any model of flow generation and propagation.
For example, in a series of recent papers by our group, differential Doppler shifts
at the borders in a number of T\,Tauri jets were reported,
suggestive of rotation around the symmetry axis 
(\citealp{Bacciotti02}; \citealp{Woitas05}; \citealp{Coffey04}; 2007).
The combination of mass outflow rate and toroidal velocity estimates allows determination of the angular momentum transported by the flow, and thus also a comparison with the angular momentum that the associated disk has to loose in order to accrete at the observed rate. 
Indeed in one test case, namely the jet from RW\,Aur, we found that the angular momentum present in the jet is at least 60 - 70\,\% of that required to be extracted from the disk (\citealp{Woitas05}).
This is of course a result of primary importance, since it could be the first 
long-awaited validation of the popular idea that jets exist to remove the excess angular momentum
in forming system, thus providing a solution to one of the main theoretical problems
in star formation. 

In order to confirm this finding, however, we have to determine the 
mass and angular momentum outflow rate close to the star, and in a wider sample of objects.
A good starting point is to use the {\it HST}/STIS spectra that were recently 
acquired by our group in a survey to search for rotation signatures in jets 
from T\,Tauri stars. The survey included jets from five  
young stars, namely TH\,28, DG\,Tau, HH\,30, CW\,Tau and RW\,Aur. 
The spectra, taken with the slit placed at a few tens of arcseconds from the star,
and oriented transversely to the flow direction, sampled the collimation region 
of the jets and allowed us to confirm the presence of differential Doppler shifts 
at the jet borders and derive estimates of toroidal velocities (\citealp{Coffey04}; 2007).
These spectra, taken with the slit perpendicular to the jet axis, included several optical forbidden emission lines that can be used to diagnose the physical conditions of the gas. 
Thus, we have embarked in a further analysis of these data to understand how the plasma properties behave close to the star where the jet is launched, collimated and accelerated, and finally to find the mass and angular momentum outflow rate in our sample. 
We point out that estimating the mass outflow rate at the jet base is more likely to exclude spurious effects, 
such as the addition of mass from possible entrainment of ambient gas and dust or 
significant disruption of the flow, which are less likely to have a dominant influence in the early stages of jet propagation. 

The determination of the mass outflow rate requires a knowledge 
of the total gas density, which, contrary to the electron density, 
is not directly measurable from the observed lines.
One, therefore, has to use some method to derive the total density from the 
available observational data, and indeed
several methods have been developed to this aim in the recent past.
For example, from a measure of the intrinsic emission line 
luminosity one can derive the number of emitting particles of a given 
species in the observed volume, which can 
then lead to the gas density under an assumption of abundances. However, this 
method relies on an accurate prior knowledge of reddening estimates, 
excitation temperature, ionisation state of the given species, and 
filling factor, all of which bring into the calculation 
substantial uncertainties (see e.g. \citealp{Nisini05}). 
An alternative method is via a determination of 
the hydrogen ionisation fraction 
of the emitting region. A direct measure of the electron 
density can be easily obtained from the [S\,II]$\lambda\lambda$6716,6731 doublet 
(see e.g. \citealp{Osterbrock89}). Dividing the electron density by the 
ionisation fraction then closely approximates the total gas density. 
One approach to finding the ionisation fraction is to model the line ratios 
under the assumption of a definite mechanism for the gas heating. 
There is not yet, however, a general consensus on the mechanism causing 
the jet emission although, without a doubt, the observed forbidden lines are excited collisionally. 
The most widely accepted explanation is that the gas is being 
heated by internal shocks, although other possibilities include ambipolar diffusion, 
turbulent mixing-layers and compression by jet instabilities 
(see e.g. \citealp{Lavalley-Fouquet00} and references therein). On the basis that the gas is shock heated, \citet{Hartigan94} constructed a grid of planar shock models
to calculate the line ratios, and compared the results to spectra of a few stellar jets integrated 
along the beam. In this way, they were able to find the most likely value of the 
hydrogen ionisation fraction in the flows, and hence the total density, after accounting for 
the shock compression. This method, however, is model dependant, and implies heavy and 
lenghty calculations. 

Subsequently, a technique was developed whereby the  
ratios of the optical forbidden emission lines of S, N and O in the red wavelength range 
are used to infer the hydrogen ionisation and the electron temperature 
{\it without} having to assume a specific method of heating, but only assuming that the atomic 
levels are populated collisionally in the absence of ionising photons
(\citealp{Bacciotti95}; \citealp{BE99}).  This so-called 'BE technique' is based 
on the recognition that, in the conditions present in stellar jets, 
the ionisation states of oxygen and nitrogen are 
tightly correlated to that of hydrogen via charge-exchange. 
Consequently, the line ratios can be easily modeled and compared with the observed 
ratios to find the most suitable values of the gas physical quantities. 
In addition to the requirement of very little calculations, 
the technique does not require an accurate knowledge of distance or extinction. 
Furthermore, the results do not depend on any specific mechanism 
for jet formation and/or evolution, and thus can be applied to targets 
in very different conditions. 

The {\it BE technique} has been applied to jet observations of varying spectral and 
spatial resolution (\citealp{Bacciotti95}; \citealp{Bacciotti96}; \citealp{Bacciotti99}; 
\citealp{Nisini05}; \citealp{Podio06}), which lately included high angular resolution observations of the DG\,Tau, RW\,Aur, LkH$\alpha$233 jets using {\em HST}/STIS multiple-slit configuration,
(\citealp{Bacciotti02mexico}; \citealp{Melnikov08}) as well as sub-arcsecond ground-based data of the DG\,Tau and RW\,Aur jets obtained with adaptive optics (\citealp{Lavalley-Fouquet00}; \citealp{Dougados02}). Very recently, \cite{Hartigan07} used a slightly different version of the {\it BE technique} to diagnose the physical condition of the HH\,30 jet from {\em HST}/STIS `slitless' spectra. 
In all cases, the method has brought very valuable information on the excitation 
of the jet gas allowing, for instance, constraints on models for jet heating, and allowing confirmation that shocks are, as expected, the most likely cause of the thermal behaviour of the plasma. In most cases, it has also been possible to estimate the mass and momentum flux in the flow from the derived total densities. This in turn has facilitated a discourse on the dynamical relationship between optical jets and coaxial molecular flows (see, e.g. \citealp{BE99}). 
Furthermore, as mentioned above, in the test case of the RW Aur jet, using the results of the {\it BE technique} and Doppler gradient measurements, we could give an estimate of the angular momentum transported by the flow. 

For all these reasons, we decided to use the {\it BE technique} to determine the mass and angular momentum flux in our sample of jets observed at high angular resolution. 
The STIS data included all the optical lines necessary to apply this procedure, which  
has proven to be very well suited to the diagnostics of large datasets in extended jet surveys, as is the case of this paper. 
In the following, we describe the application of the method to our spectra (Section\,\ref{observationsanddataanalysis}), and the obtained results (Section\,\ref{results}). 
Given the form of our input, we present the results for the physical conditions as position-velocity 
(PV) maps, in which the electron density, the ionisation fraction, the electron temperature and 
the total density are resolved both as a function of velocity and distance from the jet axis. 
From these values, we derived estimates of the mass and angular momentum outflow rates 
of the jet close to the launch point. Our findings are discussed in Section\,\ref{discussion}. 

\section{Observations and Data Analysis}
\label{observationsanddataanalysis}

\subsection{Observations}
\label{observations}

\begin{figure*}
\begin{center}
\epsscale{0.33}
\plotone{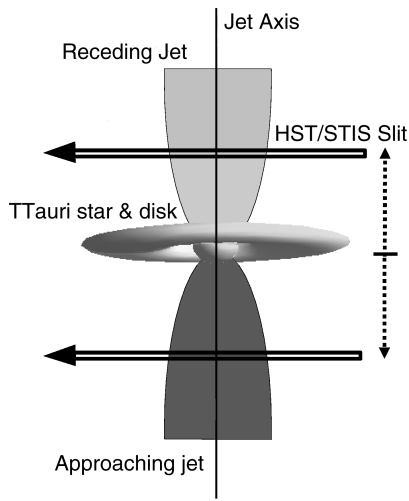}
\figcaption{Orientation of the jet and slit. The arrow on the slit indicates the positive direction of the y-axis in the position-velocity contour plots. The requested slit position angle was 90$^{\circ}$ with respect to the value of PA$_{jet}$, Table\,\ref{targets}. The actual slit position angles for DG\,Tau and HH\,30 differ from the requested values by +3$^{\circ}$ and -6$^{\circ}$ respectively. This was due to problems during observations in finding the right combination of guide stars for the requested angle. 
\label{orientation}}
\end{center}
\end{figure*}

{\em HST}/STIS spectroscopic observations at optical wavelengths (Proposal ID 9435) were made at the base of jets from several T\,Tauri stars (see Table\,\ref{targets}). These data formed the basis of two recently published papers which reported measurements of gradients in the Doppler shift of the jet radial velocity profile transverse to the flow direction (\citealp{Coffey04}; 2007). In those papers, the possibility that the measurments can be interpreted as indications of jet rotation was investigated. The adopted observing procedure involved centering the {\it HST}/STIS slit on the T\,Tauri star, rotating the slit to a position angle perpendicular to the jet axis, and then offsetting the slit to a position along the jet which is a fraction of an arcsecond from the source, as illustrated in Figure\,\ref{orientation}. 

\begin{table*}
\begin{center}
\scriptsize{\begin{tabular}{llcccccc}
\tableline\tableline
Target		&Location&Distance	&M$_{\star}$	&$v_{sys}$	&$i_{jet}$	&PA$_{jet}$ &References	\\
		&	 &(pc)		&(M${_\odot}$)	&(km\,s$^{-1}$)	&(deg)		&(deg)	&		\\ 
\tableline
TH\,28		&Lupus\,3 &170		&...		&+5		&10		&98 	&1, 2		\\ 
DG\,Tau		&Taurus	&140		&0.67		&+16.5		&52		&226 	&3, 4		\\
CW\,Tau		&Taurus	&140		&1.4		&+14.5		&41		&155 	&5, 6, 7	\\
HH\,30		&Taurus	&140		&0.45		&+21.5		&1		&33 	&8, 9, 10	\\
RW\,Aur		&Auriga &140 		&1		&+23		&44		&130   	&11, 12 	\\
\tableline
\end{tabular}}
\end{center}
\caption{Details of T\,Tauri jet targets investigated in this paper. All radial velocity results, $v_{rad}$, are quoted after correction for the systemic heliocentric radial velocity, $v_{sys}$. The inclination angle of the jet, $i_{jet}$, is 
given with respect to the plane of the sky. Values for the jet position angle, PA$_{jet}$, were determined from archival {\em HST} images. For our observations, we requested a slit position angle of 90$^{\circ}$ with respect to PA$_{jet}$, Figure\,\ref{orientation}. References - 
(1) \citealp{Graham88}; (2) \citealp{Krautter86}; 
(3) \citealp{Eisloffel98}; (4) \citealp{Bacciotti02}; 
(5) \citealp{GomezdeCastro93}; (6) \citealp{Hartmann86}; (7) \citealp{Hartigan04}; 
(8) \citealp{Pety06}; (9) \citealp{Appenzeller05}; (10) \citealp{Mundt90}; 
(11) \citealp{Woitas01},\,2002; (12) \citealp{Martin03}. 
\label{targets}}
\end{table*}

The jets were observed with slit offsets of 0$\farcs$3 in the cases of TH\,28, DG\,Tau, RW\,Aur, but 0$\farcs$2 for CW\,Tau (due to known lack of line emission at 0$\farcs$3 \citep{Woitas02}), and 0$\farcs$6 for HH\,30. (The larger offset for HH\,30 was deemed more appropriate since the star and the jet base are obscured by the system's edge-on disk.) These angular distances correspond to deprojected distances along each jet of 52, 68, 58, 37 and 84\,AU respectively. The CCD detector was used with the G750M grating, centred on 6581\,\AA, and a slit of aperture 52$\times$0.1\,arcsec${^2}$. Spectral sampling was 0.554\,\AA\,pixel$^{-1}$, corresponding to a radial velocity sampling of $\sim$25\,km\,s$^{-1}$\,pixel$^{-1}$, and spatial sampling was 0.$\arcsec$05\,pixel$^{-1}$. Resolution is twice the sampling. 

Long exposures were made of individual jet targets, each of $\sim$2000\,s~to~2700\,s (for further details see \citealp{Coffey04}; 2007). The data were calibrated through the standard {\em HST} pipeline, subtraction of any reflected stellar continuum was performed, and signals from defective pixels were removed. 

In total, this yielded eight transverse jet spectra, which included H$\alpha$, [\ion{O}{1}]$\lambda\lambda$6300,6363, [\ion{N}{2}]$\lambda\lambda$6548,6583 and [\ion{S}{2}]$\lambda\lambda$6716,6731. Data analysis is not reported for the faintest targets, namely the approaching jets from TH\,28, RW\,Aur and LkH$\alpha$\,321, due to lack of sufficient signal-to-noise for the [\ion{N}{2}] emission, which is fundamental to the diagnostic procedure used here. 

\subsection{Data Analysis}

\subsubsection{The BE Technique} 
\label{BEtechnique} 

An indepth description of the technique can be found in several 
previously published papers on the method  
and its application (\citealp{BE99}; \citealp{Podio06}; \citealp{Melnikov08}). 
Briefly, the {\it BE technique} exploits the fact that in low excitation conditions, 
such that the plasma is partially neutral and provided that no strong sources
of energetic photons are present, 
the ionisation state of oxygen and nitrogen is dominated by charge exchange with hydrogen. (Justification for this assumption was subsequently 
confirmed by \citet{Hartigan07}.) 
In such a plasma, all the sulfur can be considered singly ionised. 
Young stellar jets satisfy these conditions 
(if except is made for the so-called irradiated jets \citealp{Bally07}), because 
they have low electron temperatures and effectively no 
photoionisation, as the stellar source is usually a low mass star with 
little production of ionising radiation. Under these assumptions, the ionisation 
state of both oxygen and nitrogen can be expressed as 
a function of the hydrogen ionisation fraction and the electron temperature. 
Combining this with the fact that the [\ion{S}{2}]$\lambda$6731/$\lambda$6716 
is a well-known increasing function of the electron density 
(until it reaches the critical density of [\ion{S}{2}], i.e. $\sim$\,2~10$^{4}$\,cm$^{-3}$ ) 
means that the intensity ratio of any two of these observed emission lines 
is a {\em known} function of electron density, hydrogen ionisation fraction 
and electron temperature. In practice, the {\it BE technique} consists 
in calculating a series of model emission line ratios, and in comparing the result with the observed ratios, 
to determine the most suitable values of the physical quantities in the emitting gas. In particular, the sulfur doublet ratio is used to determine the electron density, 
while the ionisation fraction and the electron 
temperature are found by combining the ratio of nitrogen to oxygen lines, [\ion{N}{2}]/[\ion{O}{1}] 
and the ratio of the oxygen to sulfur lines, [\ion{O}{1}]/[\ion{S}{2}], at the determined electron density ratio. In fact, the first ratio is predominantly sensitive to the ionisation fraction 
([\ion{N}{2}]/[\ion{O}{1}] increases with the ionisation fraction), while the second 
is comparably sensitive to both ionisation fraction and electron temperature 
([\ion{O}{1}]/[\ion{S}{2}] increases with increasing electron temperature, 
and with decreasing electron density, for densities above the critical density). 
Once the electron density and ionisation are determined, it is possible to calculate 
the total hydrogen density in the emitting gas. 

The application of the technique is straightforward, but it had to be automated 
for the treatment of large datasets. This has been done recently upon the  
analysis of {\it HST}/STIS spectra of the LkH$\alpha$233 jet acquired by 
our group. A numerical code was written (hereafter the {\it BE code}) 
which algebraically inverts the functions 
generated by the line ratios. The code was succesfully tested on the LkH$\alpha$233 data, and the  
results were published by \citet{Melnikov08}.  

We applied this same code to our jet spectra, 
but we have updated some aspects to account for the fact that 
our data is in a higher excitation regime, i.e. close to the jet launch point. 
To compute the fractional population in the atomic levels we use a 5-level
code by \citet{Raga92} (see \citet{Bacciotti95} Appendix A for
details). We use solar abundances from \citet{Osterbrock94}. 
We update the collision strengths, partially following \citet{Hartigan07}: 
$\Omega$(\ion{S}{2}) from \citet{Keenan96}; 
$\Omega$(\ion{N}{2}) from \citet{Hudson05}; 
$\Omega$(\ion{O}{1}) from \citet{Mendoza83}. 
All the other atomic parameters (collisional ionisation rates, recombination
rates, dielectronic recombination rates, charge exchange rates) are those specified in \citet{Bacciotti99}. 

Before calculating the line ratios, a spurious shift 
in space generated by the instrument optics 
was determined for each line and removed 
(see \citealp{Coffey04}; 2007 for full details). 
Emission lines were resampled in velocity so that differences in dispersion were accounted for, 
and so ratios of emission at the same velocity could be obtained. 
We considered the brightest components in the [\ion{O}{1}] and [\ion{N}{2}] emission line 
doublets, the other being always one third of the brightest component as dictated by atomic physics. 

In calculating the electron density from the [\ion{S}{2}] doublet, values of the [\ion{S}{2}]$\lambda$6731/[\ion{S}{2}]$\lambda$6716 ratio reaching 2.15 indicate 
saturation of the doublet ratio such that it is not sensitive to higher densities. 
These positions were thus given a lower limit on electron density of 2.5\,$10^{4}$\,cm$^{-3}$
determined by the [\ion{S}{2}] critical density. Meanwhile, values of the same ratio barely reaching 0.69 cross 
the lower limit in density sensitivity, and these positions were given an upper limit 
on electron density of 50\,cm$^{-3}$. 

In applying the {\em BE code} to our jet spectra, we obtained results for electron density, 
electron temperature and ionisation fraction. The electron density was then divided by the ionisation fraction to yield an  approximation of the total hydrogen density. 

\subsubsection{Improvements on the BE code} 
\label{BE-lite} 

While each jet was observed to emit in all forbidden emission lines, it was found that the code does not converge 
at certain pixel positions. The problem was found to arise in cases where jet material is travelling 
with a broad range of velocities, and the various lines peak at different velocity components, 
i.e. specifically for the DG\,Tau jet. 
The [\ion{S}{2}] emission is more intense at lower velocities. 
Therefore, at higher velocities, the [\ion{S}{2}]/[\ion{O}{1}] ratio is often too low 
to allow convergence to a single value of the algebraic inversion performed by the {\it BE code}. 

In an attempt to overcome this problem, therefore, we eased the constraints on the algebraic calculations. 
We take the electron density derived from the [\ion{S}{2}] doublet, as before. Then, instead of using 
the {\it BE code}, we predict [\ion{N}{2}]/[\ion{O}{1}] and [\ion{O}{1}]/[\ion{S}{2}] ratios from a 
mesh of ionisation and temperature values. Comparing the predicted and observed ratios, we 
find the point where the sum of the differences squared is at a minimum. 
This directs us to which combination of temperature and ionisation values best matches 
the observed line ratios. This method of using lighter constraints, which we will call the 
{\it BE-lite code}, produced output which was confirmed to be exactly the same as 
the output from the original {\it BE code}, when applied to a jet with a 
narrow range of velocities, i.e. the TH\,28 jet. 

The {\it BE-lite code} was then applied to DG\,Tau. The original {\it BE code} was found to represent 
only 40$\%$ of the output of the {\it BE-lite code} in the analysis of the DG\,Tau jet. 
The mesh covered a temperature range from 0.2\,10$^{4}$\,K to 3\,10$^{4}$\,K in steps of 400\,K, 
and an ionisation range from 0.005 to 0.5 in steps of 0.005. 

Upon final application of the appropriate code to each target, 
the analysis of TH\,28 and DG\,Tau gave the best results. 
This is because these data provide high signal-to-noise in all emission lines. 
The analysis of the other targets was more limited, due   
to insufficient signal-to-noise in [\ion{N}{2}] emission, 
which is less than or equal to that of the other lines in all cases. 

\section{Results}
\label{results}

All targets examined were found to emit in the forbidden emission line doublets 
[\ion{O}{1}]$\lambda\lambda$6300,6363, [\ion{N}{2}]$\lambda\lambda$6548,6583 and 
[\ion{S}{2}]$\lambda\lambda$6716,6731. In the adopted observing mode, the slit does not collect light from 
the star. In fact, the slit (of width 0\farcs1) is placed perpendicular to the jet usually at 0\farcs3 from the source, and the instrument spatial line spread function has a half-width at zero-maximum of 0\farcs15 in the optical regime. 
Therefore, we can be confident that the observed emission lines originate in jet material and are not 
contaminated by stellar emission. 

\begin{figure*}
\begin{center}
\epsscale{1.0}
\plotone{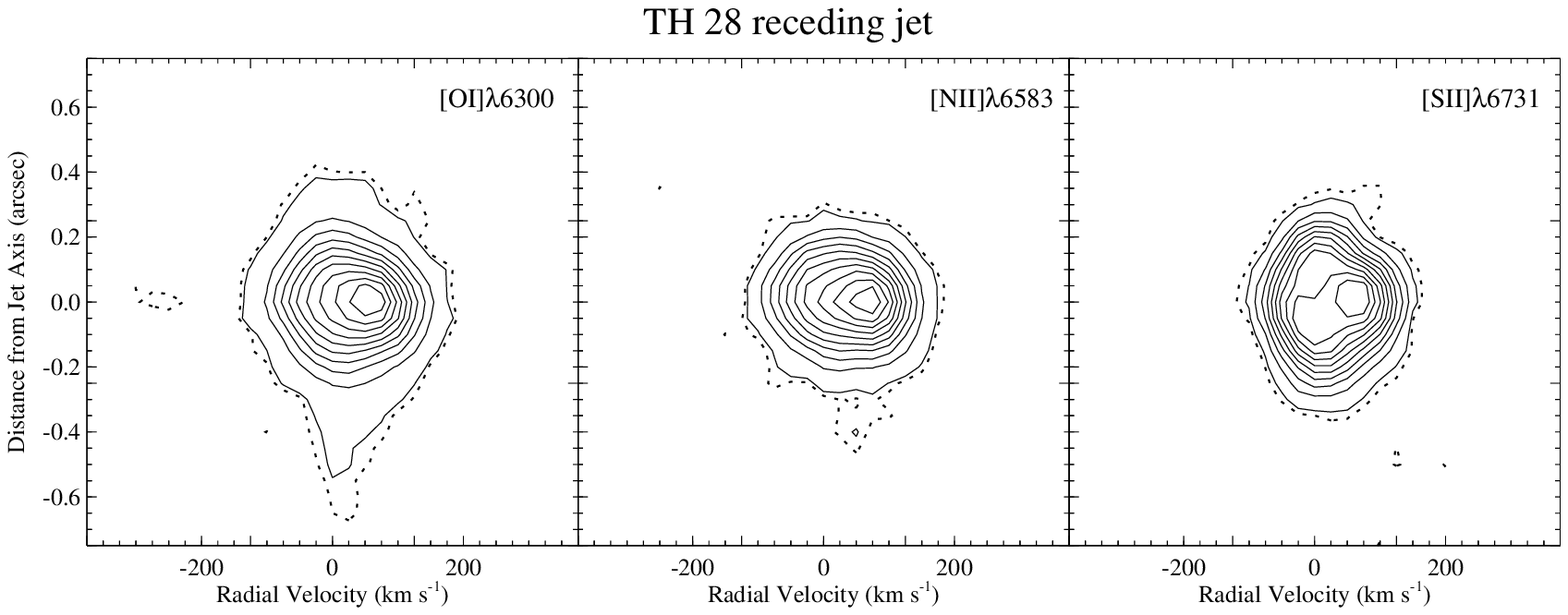}
\figcaption{Position-velocity diagrams for the brightest component in each of the three forbidden line doublets, for the TH\,28 receding jet. 
Contours begin at the three-sigma level of 1.35~10$^{-15}$ and are in steps of 10\,$\%$ up to the peak levels of 6.36, 3.55 and 2.01~10$^{-14}$\,erg\,cm$^{-2}$\,s$^{-1}$\,\AA$^{-1}$\,arcsec$^{2}$ for each plot respectively, while the two-sigma level is marked by the dotted line. 
\label{th28red_pvs}}
\end{center}

\begin{center}
\epsscale{1.0}
\plotone{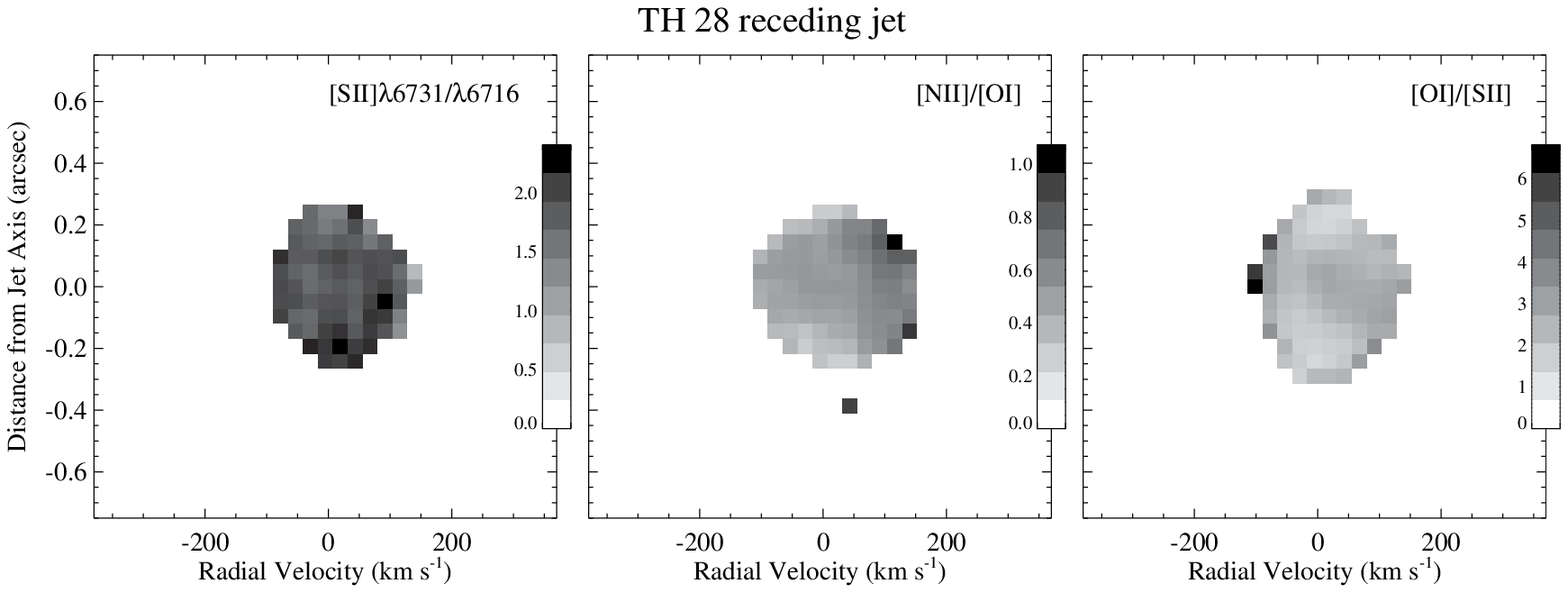}
\figcaption{Position-velocity flux ratio images demonstrating the useable points 
in each ratio, for the TH\,28 receding jet. The plots are calculated based on only 
the brightest line in the [\ion{O}{1}] and [\ion{N}{2}] doublets, from which the second component of the 
doublet was inferred using atomic physics. Positions of zero value indicate pixels with emission flux below three sigma for either one or both species in the ratio. 
\label{th28red_ratios}}
\end{center}
\end{figure*}

\begin{figure*}
\begin{center}
\epsscale{1.0}
\plotone{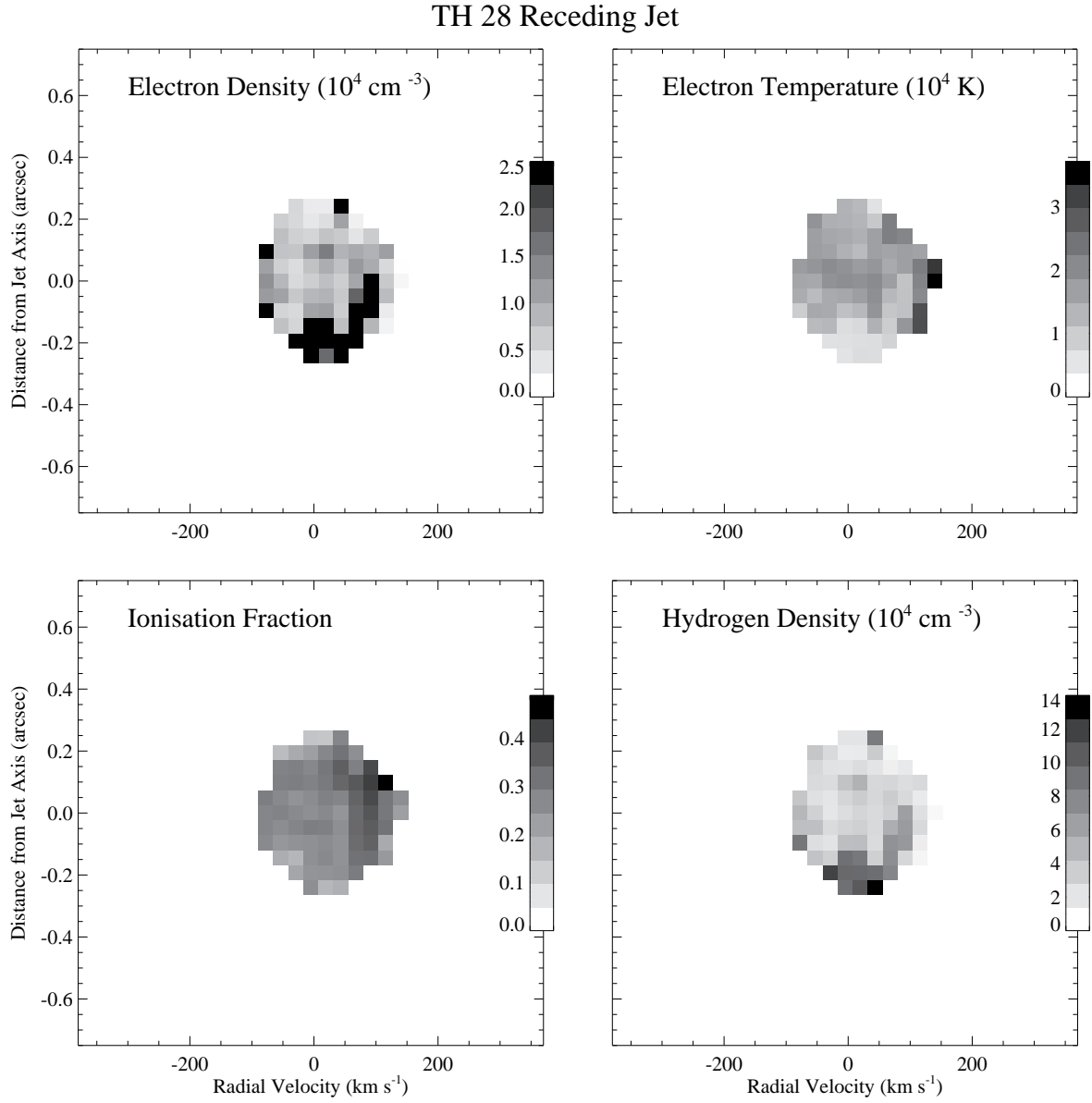}
\figcaption{
Position-velocity images showing physical jet parameters as a function of distance from the jet axis and velocity, for the TH\,28 receding jet. The plots represent electron density, electron temperature, ionisation fraction and hydrogen density calculated for each pixel of the data. 
\label{th28red_diagn}}
\end{center}
\end{figure*}

\begin{figure*}
\begin{center}
\epsscale{1.0}
\plotone{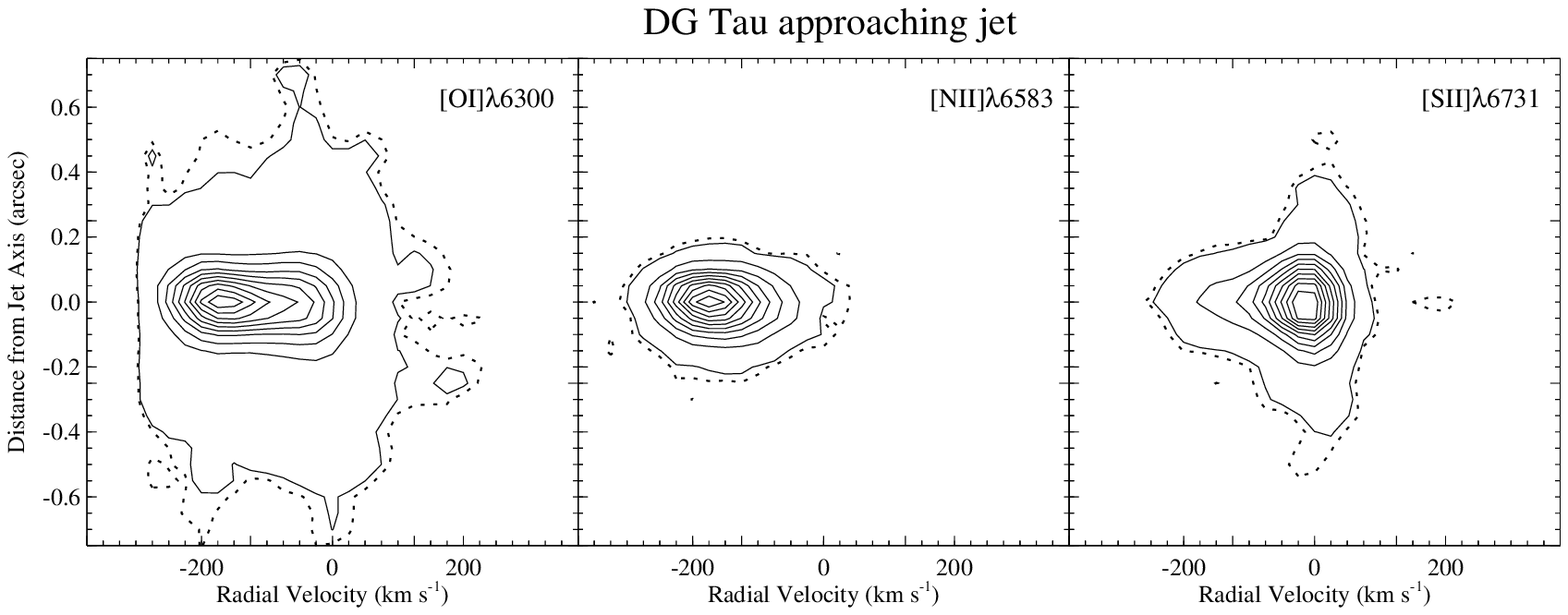}
\vspace{-1.0 cm}
\figcaption{Same as Figure\,\ref{th28red_pvs} but for the DG\,Tau approaching jet. 
Contour floor is 2.7~10$^{-15}$ and ceilings are 42.8, 7.9 and 15.5~10$^{-14}$\,erg\,cm$^{-2}$\,s$^{-1}$\,\AA$^{-1}$\,arcsec$^{2}$. 
Note that the [\ion{O}{1}]$\lambda$6300 emission is blue-shifted to the edge of the CCD. 
Therefore, for this emission line, we do not detect emission at velocities beyond -312\,km\,s$^{-1}$ for instrumental reasons. 
\label{dgtau_pvs}}
\end{center}
\vspace{-1.0 cm}

\begin{center}
\epsscale{1.0}
\plotone{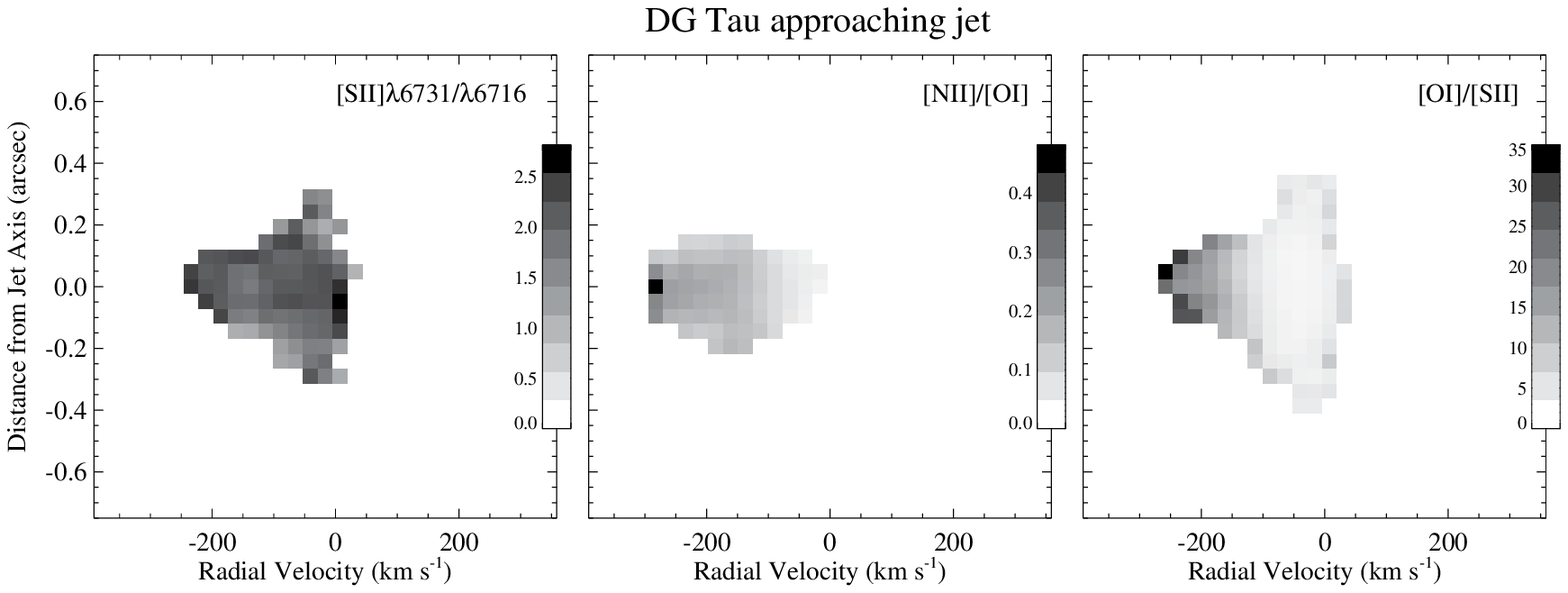}
\vspace{-1.0 cm}
\figcaption{Same as  Figure\,\ref{th28red_ratios} but for the DG\,Tau approaching jet. 
\label{dgtau_ratios}}
\end{center}
\vspace{-1.0 cm}
\end{figure*}

\begin{figure*}
\begin{center}
\epsscale{1.0}
\plotone{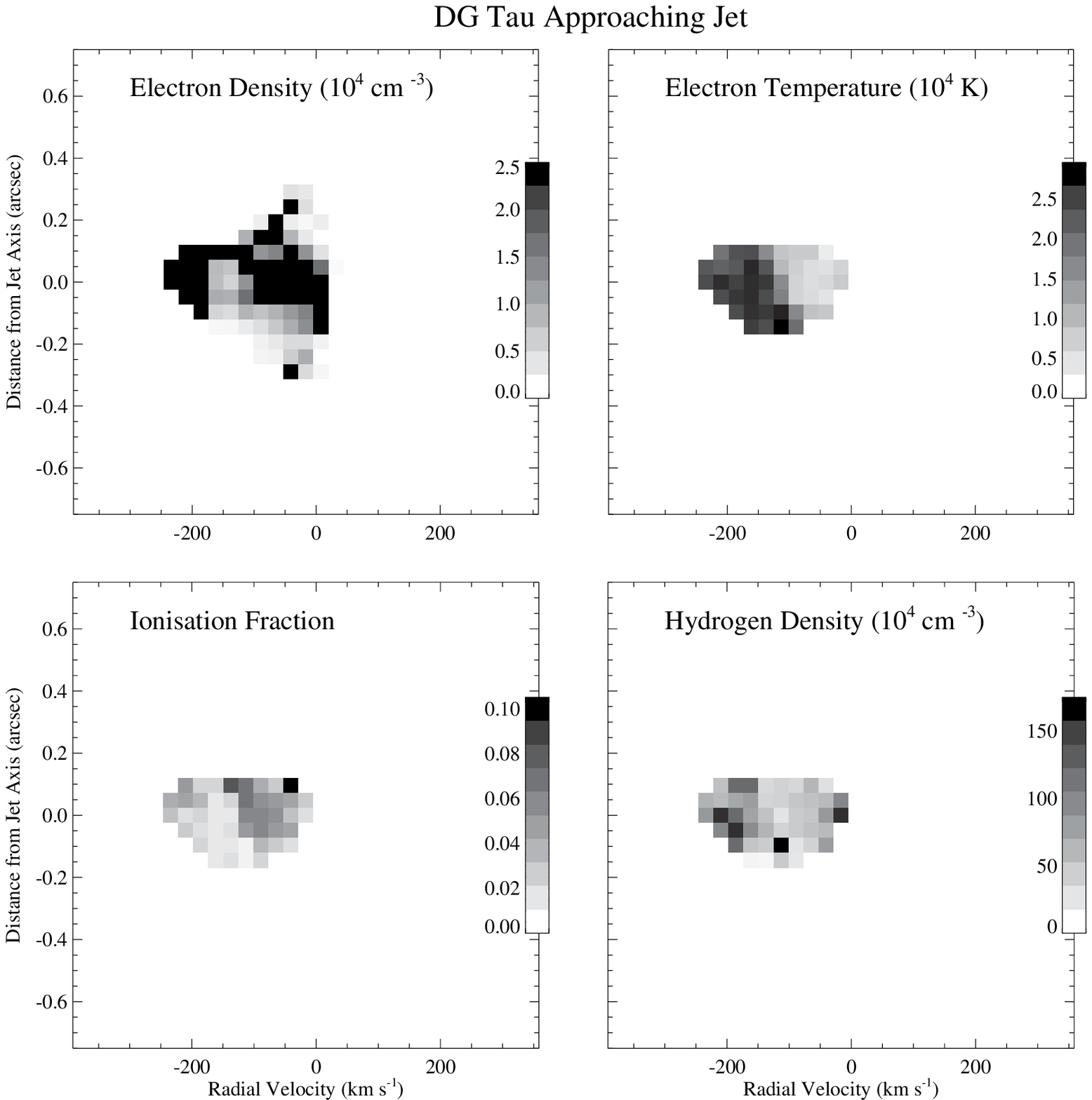}
\figcaption{Same as  Figure\,\ref{th28red_diagn} but for the DG\,Tau approaching jet. 
\label{dgtau_diagn}}
\end{center}
\end{figure*}

\begin{figure*}
\begin{center}
\epsscale{1.0}
\plotone{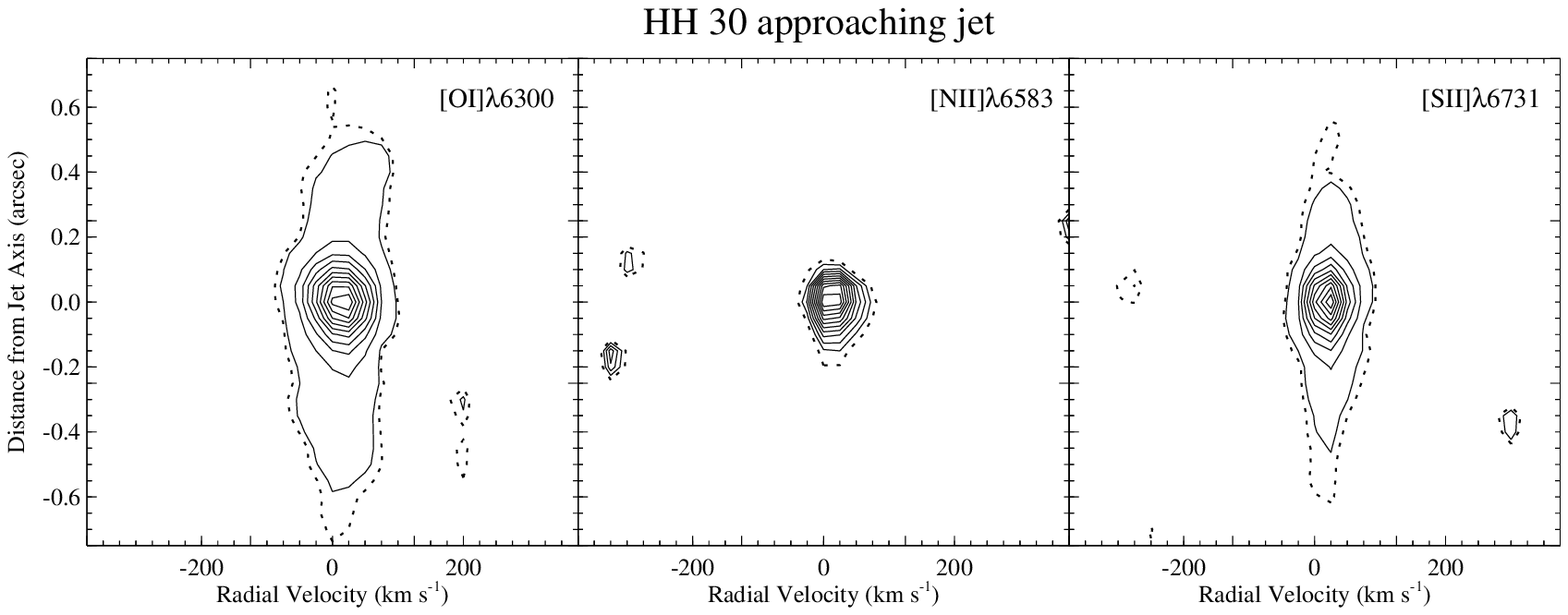}
\vspace{-1.0 cm}
\figcaption{Same as Figure\,\ref{th28red_pvs} but for the HH\,30 approaching jet. 
Contour floor is 2.4~10$^{-15}$ and ceilings are 10.9, 1.7 and 9.2~10$^{-14}$\,erg\,cm$^{-2}$\,s$^{-1}$\,\AA$^{-1}$\,arcsec$^{2}$. 
\label{hh30_pvs}}
\end{center}
\vspace{-1.0 cm}

\begin{center}
\epsscale{1.0}
\plotone{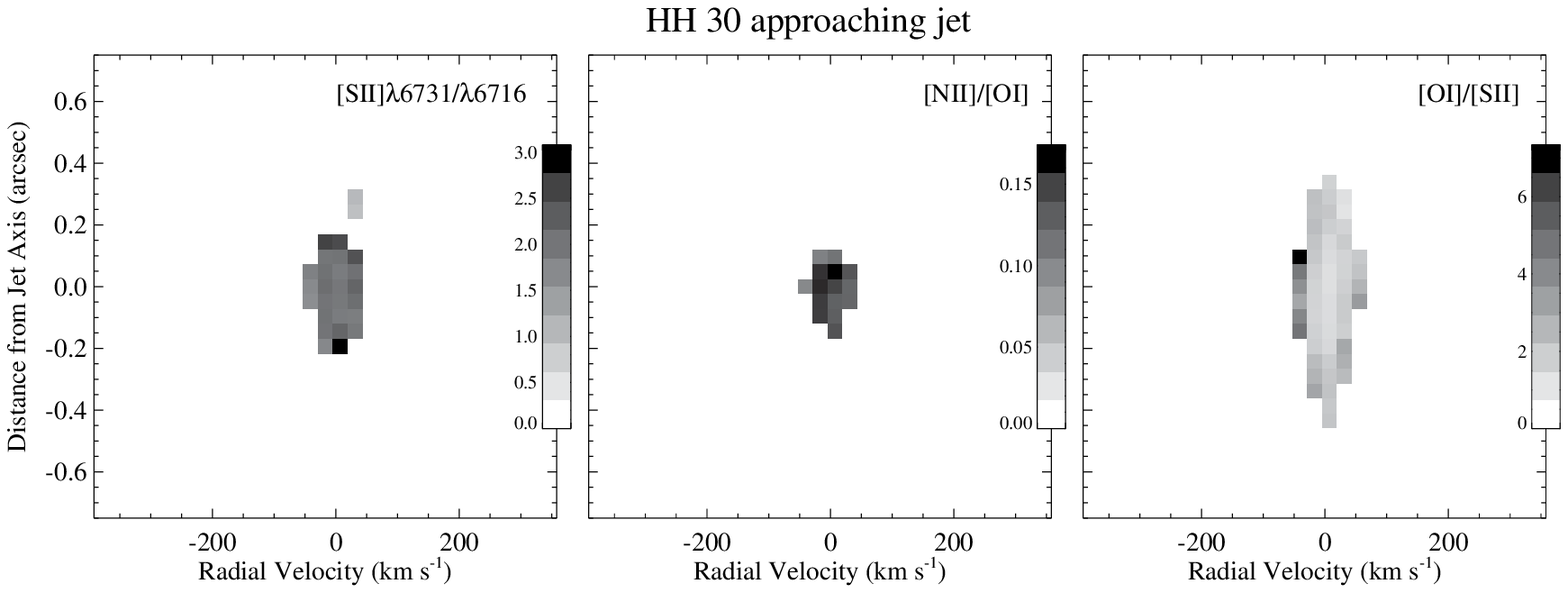}
\vspace{-1.0 cm}
\figcaption{Same as Figure\,\ref{th28red_ratios} but for the HH\,30 approaching jet. 
\label{hh30_ratios}}
\end{center}
\vspace{-1.0 cm}
\end{figure*}

\begin{figure*}
\begin{center}
\epsscale{1.0}
\plotone{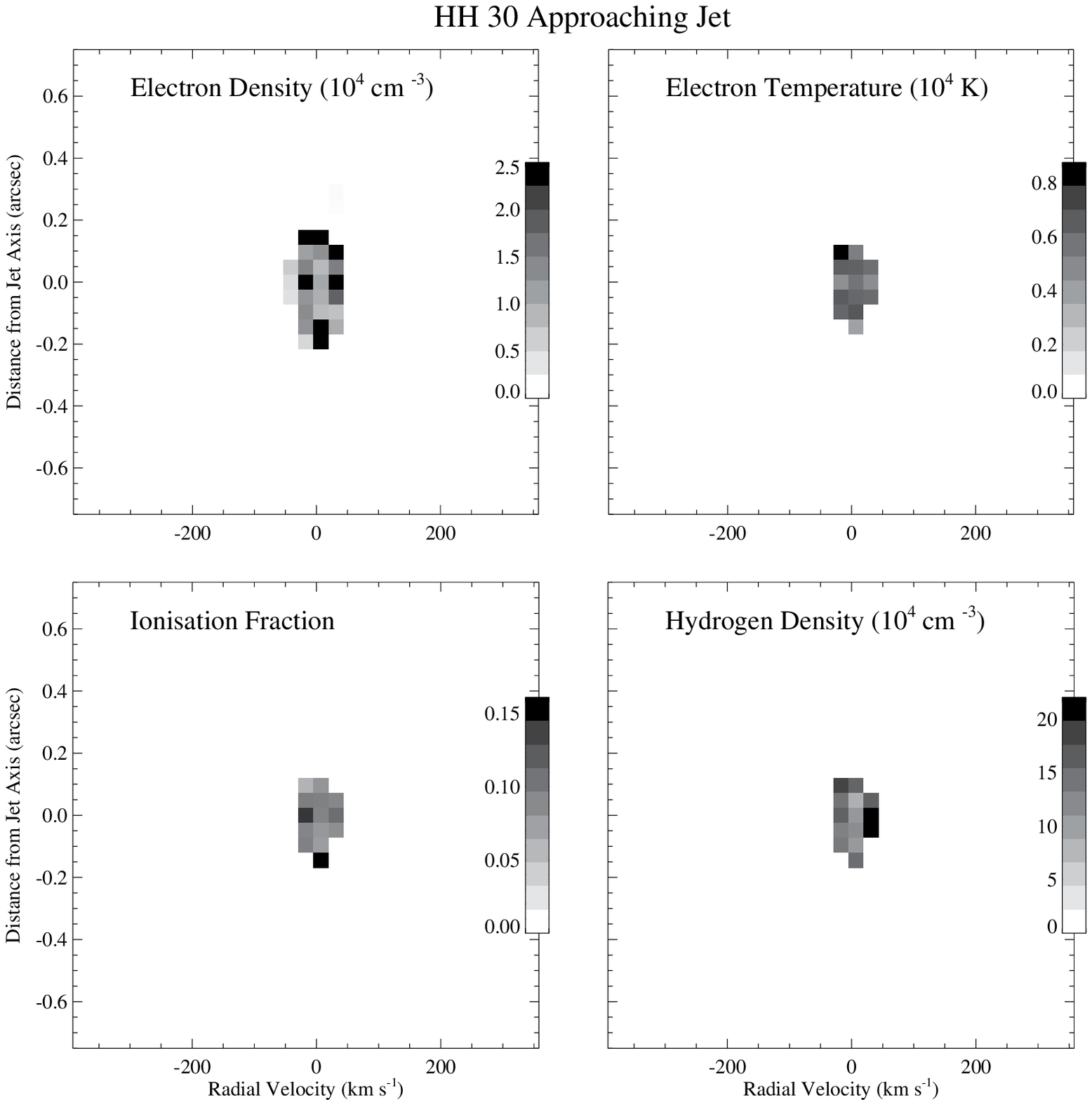}
\vspace{-1.0 cm}
\figcaption{Same as Figure\,\ref{th28red_diagn} but for the HH\,30 approaching jet. 
\label{hh30_diagn}}
\end{center}
\end{figure*}

\begin{figure*}
\begin{center}
\epsscale{1.0}
\plotone{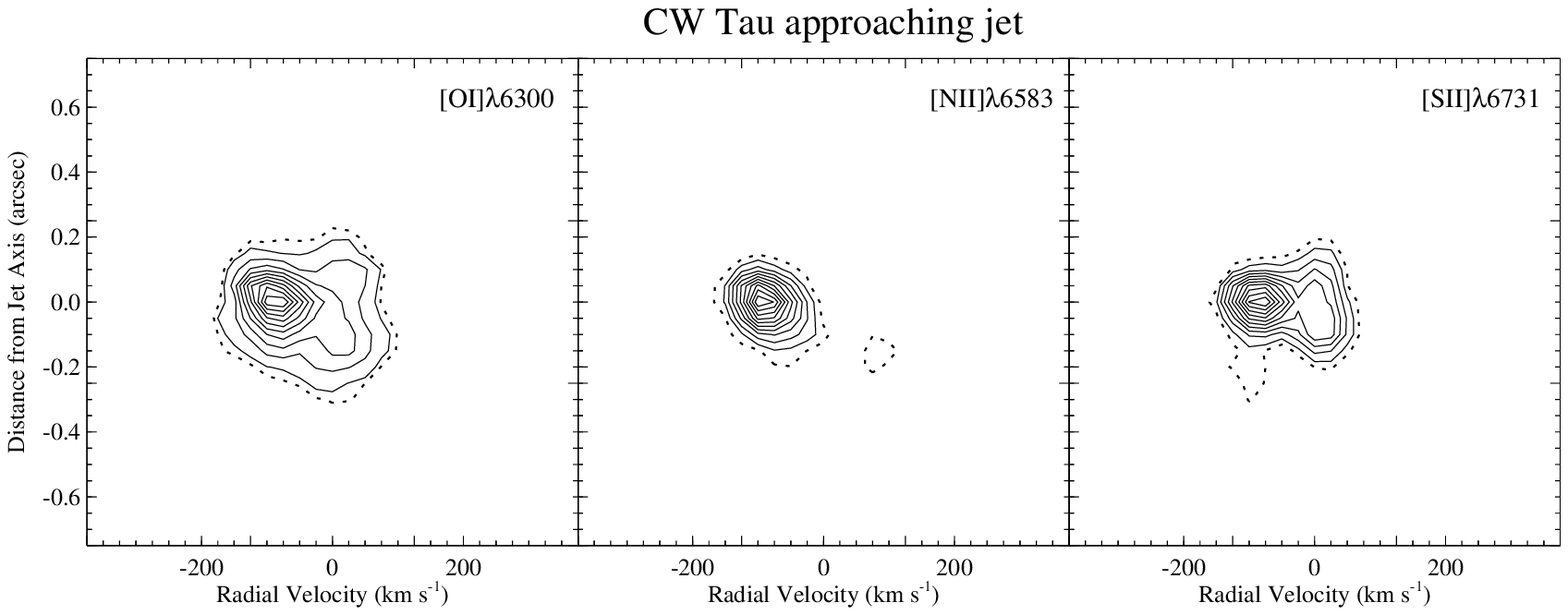}
\vspace{-1.0 cm}
\figcaption{Same as Figure\,\ref{th28red_pvs} but for the CW\,Tau approaching jet. 
Contour floor is 2.10~10$^{-15}$ and ceilings are 2.8, 1.2 and 1.2~10$^{-14}$\,erg\,cm$^{-2}$\,s$^{-1}$\,\AA$^{-1}$\,arcsec$^{2}$. 
\label{cwtau_pvs}}
\end{center}
\vspace{-1.0 cm}

\begin{center}
\epsscale{1.0}
\plotone{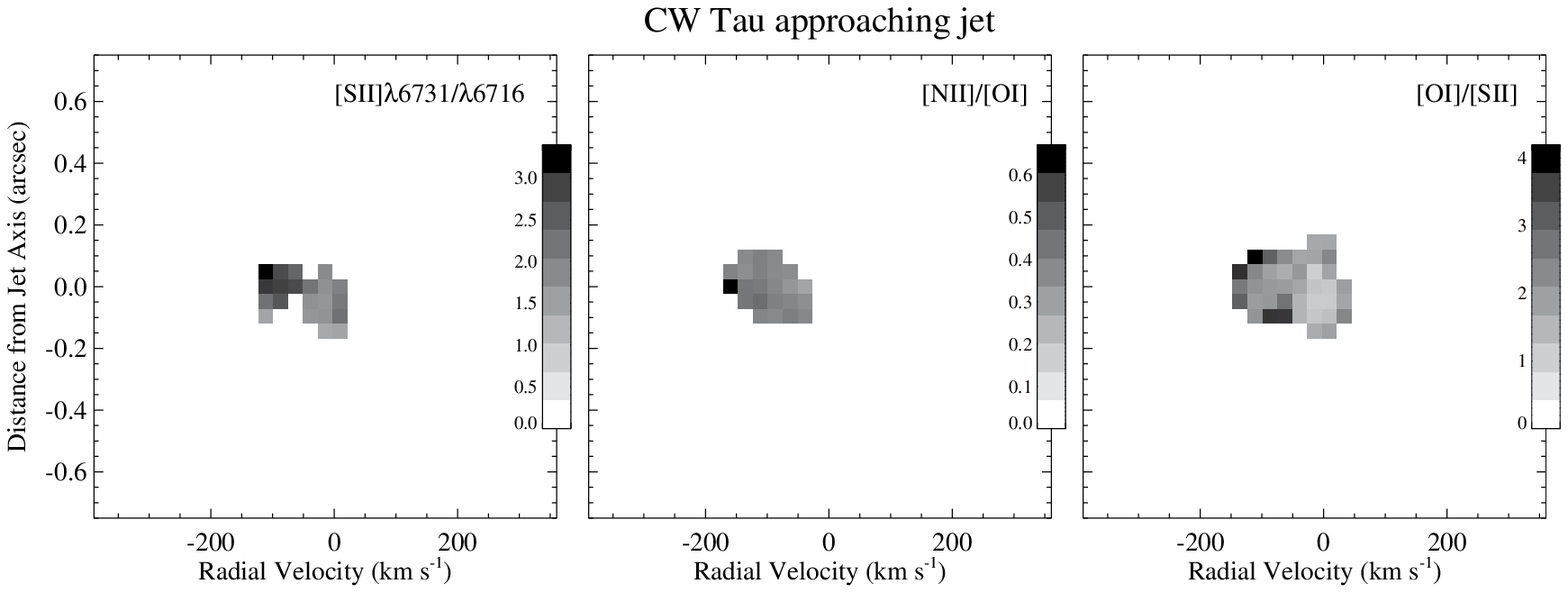}
\vspace{-1.0 cm}
\figcaption{Same as Figure\,\ref{th28red_ratios} but for the CW\,Tau approaching jet. 
\label{cwtau_ratios}}
\end{center}
\vspace{-1.0 cm}
\end{figure*}

\begin{figure*}
\begin{center}
\epsscale{1.0}
\plotone{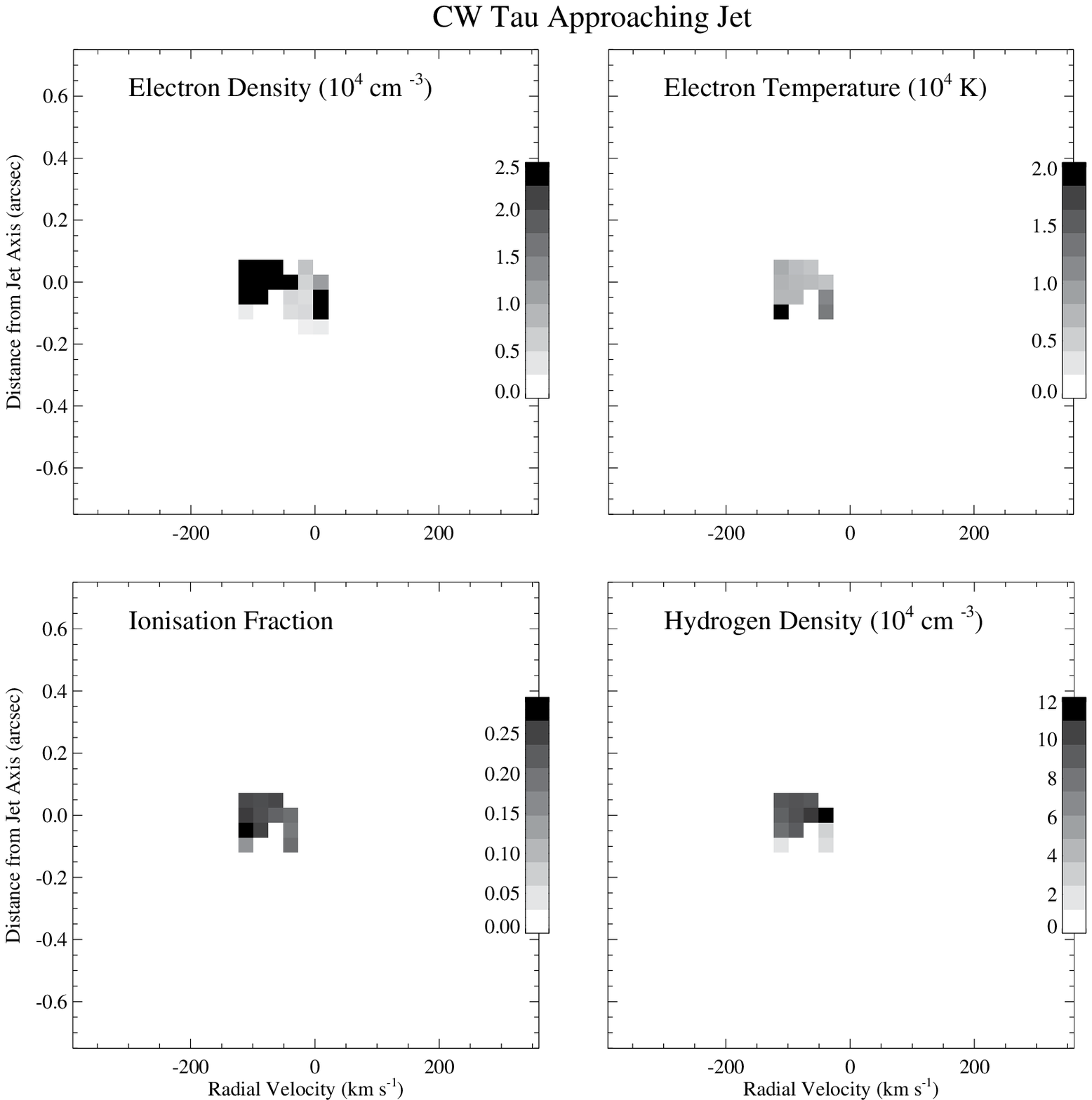}
\vspace{-1.0 cm}
\figcaption{Same as Figure\,\ref{hh30_diagn} but for the CW\,Tau approaching jet. 
\label{cwtau_diagn}}
\end{center}
\end{figure*}

\begin{figure*}
\begin{center}
\epsscale{1.0}
\plotone{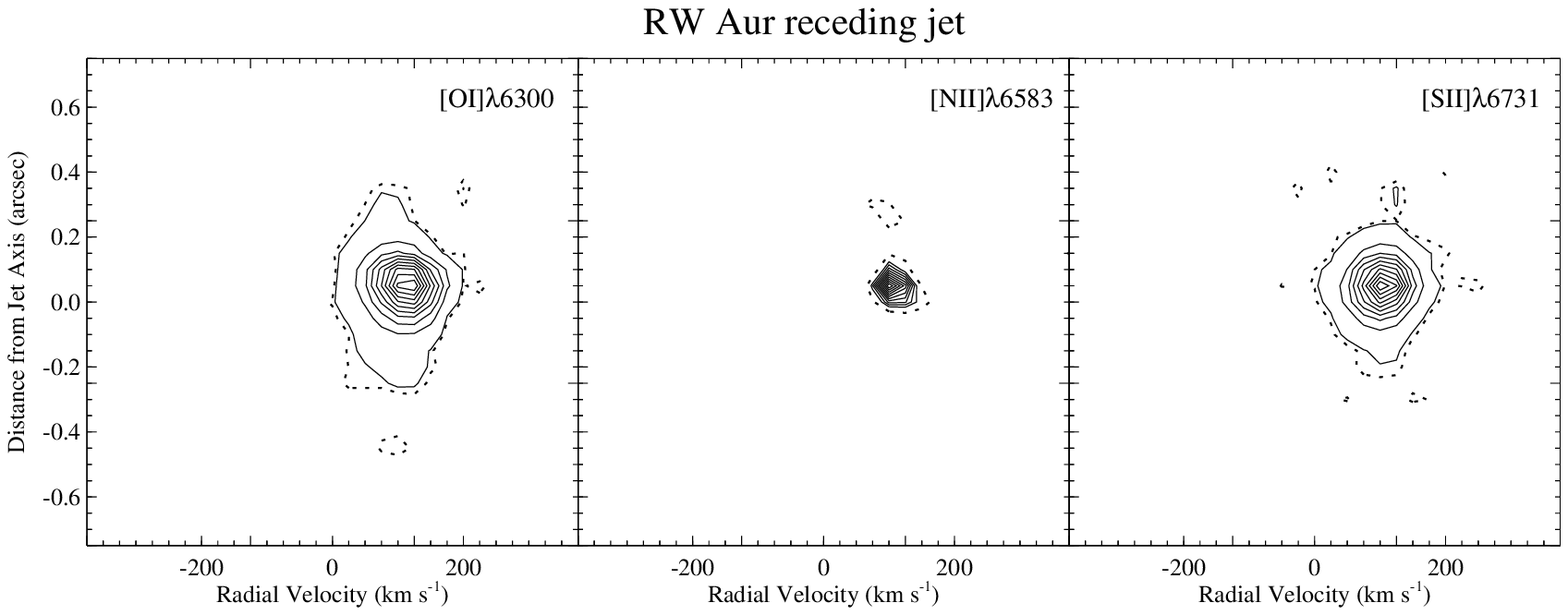}
\vspace{-1.0 cm}
\figcaption{Same as Figure\,\ref{th28red_pvs} but for the RW\,Aur receding jet. 
Contour floor is 3.0~10$^{-15}$ and ceilings are 16.3, 0.8 and 1.8~10$^{-14}$\,erg\,cm$^{-2}$\,s$^{-1}$\,\AA$^{-1}$\,arcsec$^{2}$. 
\label{rwaurred_pvs}}
\end{center}
\vspace{-1.0 cm}

\begin{center}
\epsscale{1.0}
\plotone{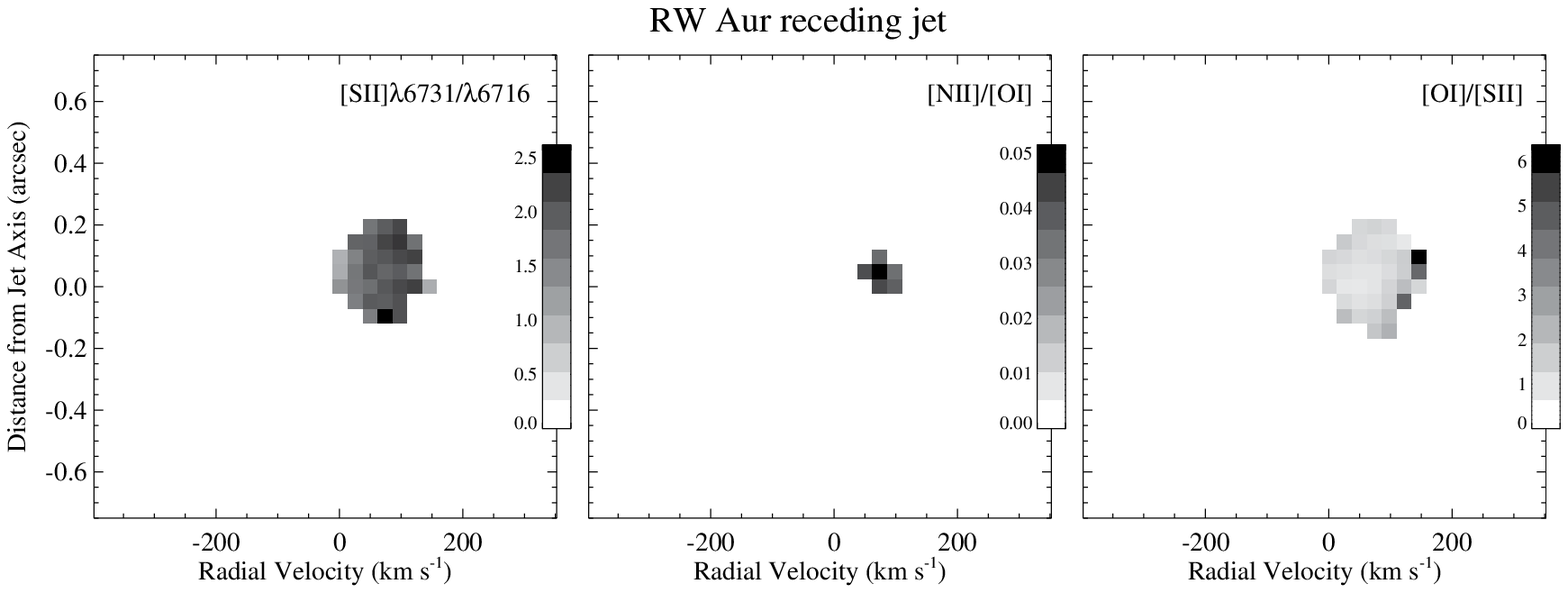}
\vspace{-1.0 cm}
\figcaption{Same as Figure\,\ref{th28red_ratios} but for the RW\,Aur receding jet. 
\label{rwaurred_ratios}}
\end{center}
\vspace{-1.0 cm}
\end{figure*}

\begin{figure*}
\begin{center}
\epsscale{1.0}
\plotone{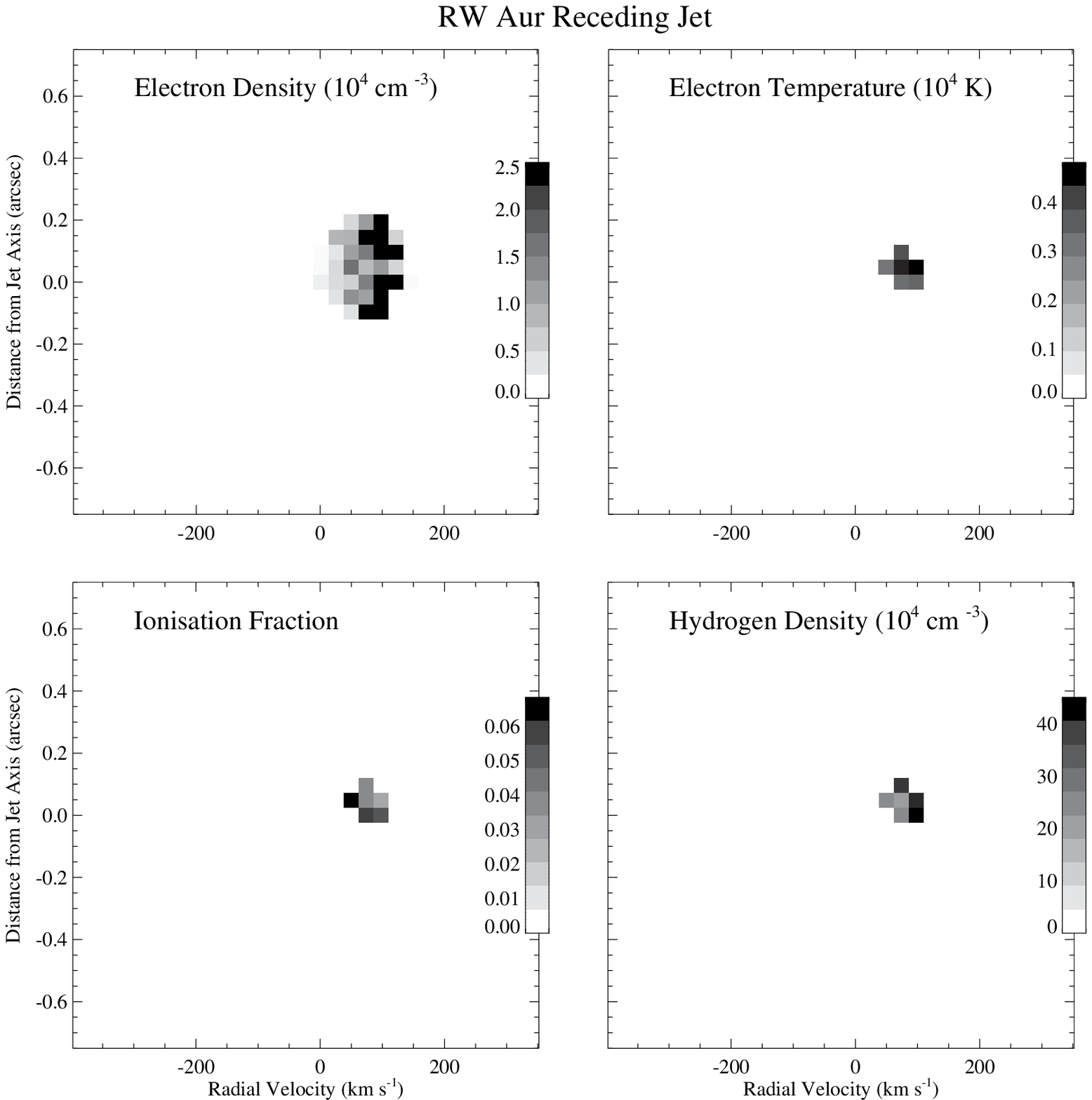}
\vspace{-1.0 cm}
\figcaption{Same as Figure\,\ref{hh30_diagn} but for the RW\,Aur receding jet. 
\label{rwaurred_diagn}}
\end{center}
\end{figure*}

Position-velocity contour maps of the brightest emission line in each doublet are shown in Figures~\ref{th28red_pvs}, \ref{dgtau_pvs}, \ref{hh30_pvs}, \ref{cwtau_pvs}, \ref{rwaurred_pvs}. Position-velocity images of 
the ratios of these lines are shown as a guide to the distribution of relative fluxes, Figures~\ref{th28red_ratios}, \ref{dgtau_ratios}, \ref{hh30_ratios}, \ref{cwtau_ratios}, \ref{rwaurred_ratios}. For ease of comparison, ratio maps are presented such that they scale positively with the relavant map of physical parameters. (Recall from Section\,\ref{BEtechnique} that [\ion{S}{2}]$\lambda$6731/$\lambda$6716] increases with electron density, 
[\ion{N}{2}]/[\ion{O}{1}] increases with ionisation fraction, 
[\ion{O}{1}]/[\ion{S}{2}] increases with increasing electron temperature and with decreasing electron density.) Position-velocity images of jet physical parameters are shown in Figures~\ref{th28red_diagn}, \ref{dgtau_diagn}, \ref{hh30_diagn}, \ref{cwtau_diagn} and \ref{rwaurred_diagn}. These figures represent output from the {\it BE code} applied to TH\,28, HH\,30, CW\,Tau and RW\,Aur, and the {\it BE-lite code} applied to DG\,Tau. The typical values are summarised in Table\,\ref{diagn_table}. 

The overall results indicate that, at the base of jets of the examined T\,Tauri stars, the electron density is high enough to saturate the [\ion{S}{2}] doublet (i.e. $>$2\,10$^4$\,cm$^{-3}$). For all targets, the doublet reaches saturation in many points. Where it does not reach saturation, the electron density is always at least above 10$^4$\,cm$^{-3}$. (It should be borne in mind that ranges for the other parameters are often derived for regions of saturated [\ion{S}{2}] doublet ratios.) The typical temperature in this region of the jet is high, and always in the range 1 - 2\,10$^4$\,K (except for RW\,Aur where it only reaches 5\,10$^3$\,K, although this dataset is the poorest). The ionisation level for all targets is in the range 0.03 - 0.3, where DG\,Tau and RW\,Aur noticeably define the lower end of this range. The total hydrogen density at the jet base lies in the range 5\,10$^4$\,-\,5\,10$^5$\,cm$^{-3}$, although this is only a lower limit for saturated regions. 

Since optically thin forbidden line emission is integrated through the jet, the results represent physical conditions seen in projection. This causes a smoothing of any gradients in the results. Since such jets are shown to have an onion-like structure, i.e. parameters vary with each layer from the jet axis \citep{Bacciotti00}, the borders of the flow (i.e. regions farthest from the jet axis) are less affected by the projection effect, because the line of sight does not cut through as many "layers of the onion". 

\subsection{TH\,28 receding jet}
\label{th28red}
The bipolar jet from TH\,28 was identified by \citet{Krautter86}. This system is
almost in the plane of the sky (10$\degr$), with a low radial velocity jet ($\sim$\,30\,km\,s$^{-1}$), and a narrow range of velocities within the jet. Low resolution spectra of this  
jet have been analysed by \cite{BE99}, who find a fall off in electron density with distance along the jet already identified by \citet{Krautter86}, and values for electron density, temperature and ionisation in the region of 10$^2$\,cm$^{-3}$, 2\,10$^{4}$\,K and 0.07 - 0.61 respectively. These values pertain to a distance of 2$\arcsec$-11$\arcsec$ from the driving source. 

Of all our targets, the results for the TH\,28 receding jet present the clearest indications of the plasma physical conditions. This is due to good signal-to-noise in all forbidden emission lines, comparable jet width in all emission lines, and a similar narrow range of velocities in all lines. The position-velocity contour plots, Figure\,\ref{th28red_pvs}, show that the jet emission is not symmetric. This is abundantly evident in the [\ion{O}{1}] plot, but is also supported by fainter [\ion{N}{2}] plot. 
The analysis reveals an asymmetry also in electron density. Although there is not much spread in the jet velocity, this asymmetry is clearly evidenced by both lower and higher velocity gas. The electron density on one side is a lower limit as it has reached the critical density for the [\ion{S}{2}] doublet. Aside from the obvious asymmetry, there is no other clear trend in electron density, which lies typically in the range of 0.5-1.5~10$^4$\,cm$^{-3}$, in the region where the sulfur doublet does not indicate saturation. 
There is, however, a trend in the temperature. Values reach 3\,10$^{4}$\,K at a thin front located at the peak red-shifted velocity, and then drop to about 2\,10$^{4}$\,K at lower velocities material. The temperature drops further still moving from on-axis to the jet borders to 1\,10$^{4}$\,K. Thus, along the jet axis, a spine of higher temperatures in the region of 2\,10$^{4}$\,K is decipherable. Also noteable is the decrease in temperature with increasing electron density on one side of the jet. This inverse relationship is noticeable to a lesser degree over the rest of the jet's cross-section. The ionisation fraction reaches 0.5 for higher red-shifted velocities. Otherwise, the ionisation appears constant over the whole jet cross-section, with a typical value for this target of 0.3. The combination of electron density and ionisation fraction leads to an overall jet hydrogen density of typically 3~10$^4$\,cm$^{-3}$. The asymmetry in electron density is reflected giving a lower limit on the hydrogen density in a few positions, reaching 1.4~10$^5$\,cm$^{-3}$. 

The behaviour of jet physical parameters which we have identified in these high resolution data shows some possible characteristics typical of those predicted by resolved shock models, e.g. \citet{Hartigan94}. According to these authors, a resolved shock is represented by high temperature which decreases with distance behind the shock, a corresponding increase in ionisation, and an increasing electron density as the gas is compressed behind the shock. The magnitude of the values depends on the speed of the shock. 
In our results, we see these parameters in velocity space and so if a shock is observed we would expect to see a high temperature peak at lower red-shifted velocities. But such a trend is not clearly identified, although we so see high ionisation at higher velocities, as would be expected of a shock resolved in velocity space. We also dont see any trend of density in velocity space, but an asymmetry in the spatial direction. 
\citet{Hartigan07} note an agreement with such expectations for temperature and ionisation trends in their analysis of low spectral resolution data for HH\,30, but remark on the obvious lack of agreement between the observations and models in that the density is expected to increase in the post-shock region but is not observed to do so. 

\subsection{DG\,Tau approaching jet}
DG\,Tau was one of the first T\,Tauri stars to be associated with an optical jet \citep{Mundt83}. This is one of the brightest jets from low mass young stars, and consequently one of the best studied amoung young stellar jets. Previous studies of the gas physics, with a view to determining the heating mechanism, were conducted by \citet{Lavalley-Fouquet00} based on sub-arcsecond spectro-imaging data taken from the ground. Gas parameters were reported as a function of distance from the source and in three velocity intervals. The datapoints closest to the star are at 0$\farcs$4. Also, Bacciotti et al. (2000; 2002) reported physical parameters derived from {\it HST}/STIS data of initial portion of this jet with the slit parallel to the flow direction. 

The contour plots, Figures~\ref{dgtau_pvs}, reveal that this jet emits in two clearly identifiable velocity bins, 
the higher velocity component (HVC) and the lower velocity component (LVC). 
[\ion{N}{2}] is clearly more collimated and traces high velocities, while [\ion{S}{2}] is spatially extended and 
traces low velocities. [\ion{O}{1}] traces both LVC and HVC with its peak at high velocities, 
and demonstrates that the LVC is indeed less collimated than the HVC. 
The nominal boundary between HVC and LVC was determined to be -100\,km\,s$^{-1}$, 
by two-Gaussian deblending of the [\ion{O}{1}]$\lambda$6300 line profile. 
Note that the [\ion{O}{1}]$\lambda$6300 emission is blue-shifted to the edge of the CCD. 
Therefore, for this emission line, we do not detect emission at velocities beyond -312\,km\,s$^{-1}$ for instrumental reasons. 

As a direct result of the large velocity dispersion within this jet, 
and slightly difference velocity ranges for different lines, 
the results for DG\,Tau presented here are produced with the {\it BE-lite code}. 
Immediately, we note the saturation of the [\ion{S}{2}] doublet in many points, both in the HVC and in the LVC. Because of this saturation, it is not possible to define a more detailed trend for electron density as a function of velocity. Spatially, however, we can clearly see that the electron density is higher close to the jet axis then at the jet borders, in agreement with \citet{Bacciotti00}. A similar spatial trend in other parameters could not be established since [\ion{N}{2}] was not observed in the jet borders. A typical electron temperature of 2.5\,10$^{4}$\,K was found for the HVC and 0.5\,10$^{4}$\,K for the LVC respectively. The overall results for ionisation show the observed region of the jet, at 0$\farcs$3 from the source, to be of low ionisation with values reaching only 0.02 in the HVC and 0.05 for the LVC. 
Recall that we report the electron density in the LVC as a lower limit, 
implying that the temperature and ionisation derived at these positions should be treated with some caution,
because using the critical density instead of the true density gives high values of temperature 
and low values of ionisation. 
We then find a total hydrogen density to be quite scattered, with a tendancy for the HVC to be more dense. 
Typical values reach $\sim$\,10$^{6}\,cm^{-3}$ in the HVC and 5\,10$^{5}$\,cm$^{-3}$ in the LVC. 
Spatially, no trend in the temperature, ionisation and total density was found. 

Our results agree with trends close to the star reported in the literature. 
\citet{Bacciotti02mexico} report the same low ionisation levels in the same position 
along the jet as our observations. They also find the same trend 
in the ionisation being higher for the LVC than the HVC, in agreement with our results. 
However, they find higher electron density for the HVC than the LVC, 
whereas we find saturation in both cases. 
Finally, it is important to highlight that the high electron density 
which we find at the base of the jet is not found further along the 
jet \citep{Lavalley-Fouquet00}. 

\subsection{HH\,30 approaching jet}
\citet{Mundt83} identified this jet, upon observing emission line nebulosity in the HL Tau complex in Taurus. This jet is in the plane of the sky, with a low radial velocity ($\sim$\,-5\,km\,s$^{-1}$) and narrow range of velocity within the jet. HST images of this system \citep{Burrows96} showed for the first time a disk-jet perpendicular configuration in young stars. Previous diagnostic studies focused on this target include \citet{Bacciotti99} and \citet{Hartigan07}, both of which are based on data of high spatial resolution but low spectral resolution. 

There is no division of HVC and LVC in the case of HH\,30, since the internal velocity dispersion is small. Datapoints here are fewer, with respect to the previous two targets, due to the low level of [\ion{N}{2}] emission. HH\,30 shows a scatter in the distribution of electron density, but with all values above $\sim$10$^4$\,cm$^{-3}$. Its temperature range is scattered within the range 0.4 - 0.8\,10$^4$\,K, and its ionisation lies in the range 0.08 - 0.15. The total hydrogen density is thus typically 1.5\,10$^5$\,cm$^{-3}$.

\citet{Bacciotti99} presented a similar analysis of HH\,30, but in one dimension along the jet and low spectral resolution, showing how each parameter varies with distance from the source. Close to the source, they report an electron temperature of 2~10$^{4}$\,K, and a lower limit on electron density due to saturation of the [\ion{S}{2}] doublet ratio. They report that the jet ionisation fraction rapidly rises from 0.065 at 0$\farcs$2 to 0.1 at 0$\farcs$4, and then slowly increases up to 0.140 within 2$\arcsec$ from the source. Our results are in good agreement with these values for the examined position. 

More recently, \citet{Hartigan07} conducted a study of the HH\,30 jet using HST 'slitless' spectroscopy. This method relies on the fact that choosing a wide (2$\arcsec$) slit ensures the jet is unresolved spectrally, but produces sufficient dispersion to separate images of the jet in the different optical emission line. The data represents the jet to 4\arcsec (600\,AU) along the flow, and in two epochs. Hence, the analysis shows how jet parameters vary with distance and time. Our results are in excellent agreement with those reported at 0\farcs3 from the source in both epochs (i.e. log electron density $\sim$\,4.6, electron temperature $\sim$\,5000\,K, ionisation fraction $\sim$\,0.1). 

\subsection{CW\,Tau approaching jet}
The jet from this source was discovered in optical images by \citealp{GomezdeCastro93}. Although this jet is well studied both through high resolution imaging and spectroscopic observations (e.g. \citealp{Dougados00}; \citealp{Hartigan04}; \citealp{Coffey07}), no previous analysis of the gas physics has been published for this target. 

CW\,Tau very clearly exhibits a separation of jet emission into a HVC and LVC, with the nominal boundary between the two measured as -50\,km\,s$^{-1}$ using two-Gaussian deblending. As with HH\,30, points are few due to the level of [\ion{N}{2}] emission. From the available datapoints, both velocity components appear saturated in electron density with some unsaturated points in the LVC at typically 0.8\,10$^4$\,cm$^{-3}$. Values are found in electron temperature of 0.8\,10${^4}$\,K and ionisation was typically of 0.2 - 0.3. Total hydrogen density then became 8\,10${^4}$\,cm$^{-3}$. 

\subsection{RW\,Aur receding jet}
RW\,Aur is a complex triple star system. The highly collimated jet from RW\,Aur A was identified by \citet{Hirth94}. Previous studies of the gas physics have been conducted \citep{Dougados02}, and gas parameters reported as a function of distance from the source and in three velocity intervals. In that study, the datapoints closest to the star are at 0$\farcs$4. 

The analysis of the RW\,Aur jet yields just a few datapoints in the diagnostic maps, due to the poor [\ion{N}{2}] emission. The [\ion{S}{2}] doublet begins to saturate and leaves a scattered pattern which can be vaguely observed to increase with increasing red-shifted velocities. The derived temperature reaches 0.5\,10$^4$\,K. Like the DG\,Tau jet, RW\,Aur also indicates low ionisation at the observed location along the jet, with a maximum of 0.07. Our results are in agreement with previous studies \citep{Dougados02}, which give upper limits at the jet base to be $\sim$\,10$^4$\,K for the temperature, $\sim$\,10$^{-2}$ for the ionisation fraction, $\sim$10$^4$\,cm$^{-3}$ for the electron density. The resulting total hydrogen density reaches 5\,10$^5$\,cm$^{-3}$. 

\subsection{Errors}

Sources of error on the estimated parameters depend on the flux measurement error, the effects due to differential extinction, the choice of elemental abundances. 

The error on all fluxes is calculated as the rms of the background, after cleaning cosmic rays and defective pixels. This varies according to target, but remains relatively constant across all emission lines for a given target spectrum. From this we obtain the three sigma threshold reported as the contour floor in each contour plot of the emission lines. Errors on each of the derived physical parameters due to uncertainty on the line flux measurement were found to be on the order of 30$\%$ for points of good signal-to-noise. 

The amount of dust extinction towards the various positions along these jets is unknown, and may even vary along the beam of a single object (see \citealp{Nisini05}; \citealp{Podio06}). The extinction towards the targets in this study could not be calculated, since the emission lines observed are too close to each other in wavelength. Therefore, we have not accounted for the effects of extinction by dereddening our emission line fluxes. Differential reddening, i.e. the variation of extinction with wavelength, does affect flux ratios. However, since we are dealing with emission lines spanning a small range in wavelength, the effect is not expected to be significant. Previous studies confirm that this is indeed the case. \citet{Hartigan07} examined the effects of differential reddening on their study of HH\,30. They find that the uncertainty introduced by an extinction of A$_v$$\sim$\,1 is not significant compared to the uncertainties in the relative abundances, and hence choose to ignore the effects in their analysis. \citet{BE99} report that, assuming a fiducial extinction of A$_v$$\sim$\,3, variations in relative flux values reveal errors in the final jet parameters of amounts no larger than the measurement errors, being at most 10$\%$ for the ionisation fraction and 15$\%$ for the temperature. For each of our sources, the level of extinction is A$_v$$<$3 (\citealp{Graham88}; \citealp{Beckwith90}; \citealp{Hartigan07}; \citealp{Eiroa02}; \citealp{Ghez97}). Furthermore, the visual extinction of T\,Tauri stars is expected to sharply decrease moving away from the source, and hence should be much lower than 3 at 0$\farcs$3 - 0$\farcs$6. Therefore, we neglect the error introduced by the fact that we are not dereddening the fluxes prior to the analysis. 

Lastly, the {\it BE code} relies on an assumption of elemental abundances for the determination of temperature and ionisation. Solar abundances are adopted, but may not accurately reflect conditions in star forming clouds. \citet{Podio06} show that, using abundances determined for the interstellar medium in Orion, the values of ionisation and temperature inferred using the {\it BE technique} are within 15$\%$ of those obtained assuming the most recent determinations of solar abundances. Since our errors arising from flux measurements are in the region of 30$\%$, we determine that only an insignificant error is introduced to our results by adopting solar abundances, instead of the actual abundances for the considered star forming regions (i.e. Taurus-Aurigae and Lupus). 





\begin{table*}
\begin{center}
\scriptsize{\begin{tabular}{lcccccccc}
\tableline\tableline
Target				&$\overline{n_e}$ 	&$\overline{T_e}$	&$\overline{x_e}$	&$\overline{n(H)}$	\\
				&(10$^{4}\,$cm$^{-3}$)	&(10$^{4}$\,K)		&			&(10$^{4}\,$cm$^{-3}$)	\\ 
\tableline
TH\,28 receding jet		&1.0			&2.0			&0.30			&3	\\ 
DG\,Tau approaching jet HVC	&$>$2.5			&2.5			&0.02			&90	\\
DG\,Tau approaching jet LVC	&$>$2.5			&0.5			&0.05			&50	\\
HH\,30 approaching jet		&1.5			&0.6			&0.1			&15	\\
CW\,Tau approaching jet HVC	&$>$2.5			&0.8			&0.30			&8	\\
CW\,Tau approaching jet LVC	&0.8			&1.2			&0.20			&4	\\
RW\,Aur receding jet		&$>$2.5			&0.5			&0.06			&30	\\
\tableline
\end{tabular}}
\end{center}
\caption{Summary of typical values of the output from the {\it BE technique} analysis, as shown in detail in Figures\,\ref{th28red_diagn}, \ref{dgtau_diagn}, \ref{hh30_diagn}, \ref{cwtau_diagn} and \ref{rwaurred_diagn}. The lower limit on electron density is not carried through to the total density for reasons explained in Section\,\ref{discussion}.
\label{diagn_table}}
\end{table*}

\section{Discussion}
\label{discussion}

As discussed in Section\,\ref{introduction}, the fundamental parameter to be derived from the diagnostic analysis is the mass outflow rate of the flow. Mass flux was determined from an average of the total hydrogen density, $\overline{n(H)}$, the 
poloidal velocity of the emission peak for [\ion{O}{1}], $v_p$, and the jet radius, $r_{jet}$, in order to get an average value integrated over the full jet cross-section. The formula used was $\dot M_{jet} = \mu m_p \overline{n(H)} \pi r^2_{jet} v_p$, where $\mu$ is the mean molecular weight ($\sim$\,1.24) and $m_p$ is the mass of a proton. The jet poloidal velocity was determined from radial velocity measurements as $v_{p}={v_{rad}}/\sin\,i_{jet}$. For HH\,30, this value is taken from proper motion measurements close to the source \citep{Burrows96}, since the inclination angle is so small that poloidal velocities cannot be determined from radial velocity measurements without incorporating considerable error. The jet cross-section was determined by assuming that the FWHM of the [OI]$\lambda$6300 emission can be taken as the jet diameter, since it traces both HVC and LVC. In cases where the jet material is clearly divided into HVC and LVC, i.e. for DG\,Tau and CW\,Tau, the mass outflow rate was estimated for each component separately. The estimates are reported in Table\,\ref{massflux}. 
Since the measurement uncertainties on the quantities in the calculation are at the 30$\%$ level, 
the uncertainty on the calculation of the mass outflow rate leads to a factor of a few, similar to previously 
published estimations. We note that any mass outflow rate estimate conducted in this way 
can represent an upper limit since, not only does this method assumes a homogenous distribution of material within the jet, but also the densest regions dominate the observed emission leading to a bias towards higher densities. 
However, close to the star we do expect a homogenity and high density. Conversely, the positions where the [\ion{S}{2}] doublet saturates indicate a lower limit on the electron density and hence on the total density measurement. Therefore, considering these counter-acting factors, the mass outflow rate is more of an average indication, rather than an upper or lower limit. 

Angular momentum flux estimates were obtained by assuming the radial velocities and mass outflow rates were indicative of gas physics over the full jet cross-section. Values were determined by combining the mass outflow rate with toroidal velocities derived from the differences in radial velocity across the jet. The average toroidal velocity of the jet is calculated from the transverse radial velocity gradient (\citealp{Coffey04}; 2007) and the jet inclination angle as $\overline{v_\phi}=(\triangle v_{rad}/2)/\cos\,i_{jet}$. The radius from the jet axis was taken as half the FWHM, as measured for the mass outflow rate calculation. Since we approximate the mass outflow rate and toroidal velocity, and presume them to be unvarying over the full cross-sectional area of the jet, we are satisfied with a simplified formula for the integration of the angular momentum outflow rate, $\dot L_{jet} = \dot M_{jet} v_{\phi} r_{jet}$. 
For HH\,30, no radial velocity difference were reported above the error bars \citep{Coffey07}, and so no angular momentum outflow rate was calculated. The results are reported in Table\,\ref{massflux}. 

\subsection{TH\,28 receding jet}

For this jet, the mass and angular momentum outflow rates in this lobe are of the order of those previously reported for T\,Tauri jets, although on the low side, with values of 1.2~10$^{-8}$\,M$_{\sun}$\,yr$^{-1}$ and 2.5~10$^{-6}$\,M$_{\sun}$\,yr$^{-1}$\,AU\,km\,s$^{-1}$ respectively. 
The asymmetry in electron density identified in this target translates to a mass outflow rate which is doubled on one side of the jet axis compared with the other side. Values were calculated for two semi-circular areas (seen in projection) either side of the central pixel row which represents the jet axis. For the calculation, the poloidal velocity of the emission peak was used, along with the average hydrogen density either side of the axis. This yielded estimates of 0.4 and 0.9~10$^{-8}$\,M$_{\sun}$\,yr$^{-1}$ for the two semi-circles. Furthermore, the latter value relies heavily on regions where the electron density is high enough to cause saturation of the [\ion{S}{2}] doublet ratio and so it is likely to represent a lower limit (if the jet is homogenous at this position). 

\subsection{DG\,Tau approaching jet}

For DG\,Tau, the jet material is clearly divided into HVC and LVC, 
and so the mass outflow rate was estimated for each component separately. 
The mass outflow rate is higher in the 
HVC at 4.1~10$^{-8}$\,M$_{\sun}$\,yr$^{-1}$ 
compared to 2.6~10$^{-8}$\,M$_{\sun}$\,yr$^{-1}$ in the LVC. 
\citet{Bacciotti02} take a nominal boundary of -70\,km\,s$^{-1}$ 
between the {\em medium} velocity component and LVC (instead of -100\,km\,s$^{-1}$), 
and see the trend reversed. They find values of 5.1 and 8.3~10$^{-8}$\,M$_{\sun}$\,yr$^{-1}$ 
as determined from HST/STIS spectra with the slit placed parallel to the flow axis. 
Our total mass outflow rate of 6.7~10$^{-8}$\,M$_{\sun}$\,yr$^{-1}$ 
is somewhat lower than 1.3~10$^{-7}$\,M$_{\sun}$\,yr$^{-1}$ for the parallel slit data, 
but the latter assumed jet widths for each velocity component which were 
based on slit positions, and which transpired to be 1.5 times our FWHM measurements. 
The mass accretion rate for this system is measured as 2~10$^{-6}$\,M$_{\sun}$\,yr$^{-1}$ \citep{Hartigan95} 
and so, doubling our values to account for a bipolar flow, 
we find an mass flow ratio of $\dot M_{jet}/\dot M_{acc}$\,$\sim$\,0.07. 
Our results show that the jet angular momentum outflow rate in the approaching jet is also considerably higher in the HVC. 
We found estimates of 8.4 and 4.5~10$^{-6}$\,M$_{\sun}$\,yr$^{-1}$\,AU\,km\,s$^{-1}$ for HVC and LVC respectively. 
The total of 1.3~10$^{-5}$\,M$_{\sun}$\,yr$^{-1}$\,AU\,km\,s$^{-1}$ 
represents the highest angular momentum outflow rate of all our targets. 
The value is in good agreement with 3.8~10$^{-5}$\,M$_{\sun}$\,yr$^{-1}$\,AU\,km\,s$^{-1}$ \citep{Bacciotti02}, considering the aforementioned difference in the adopted jet radii. Our high spatial resolution results show higher values of angular momentum outflow rate closer to the axis, which is traced by the HVC. In this jet HVC and LVC are well separated and so projection effects may be lessened, given that the HVC traces the core of the jet while the LVC traces the outer borders. 

\subsection{HH\,30 approaching jet}
The mass outflow rate of HH\,30 is very low at 4.0~10$^{-9}$\,M$_{\sun}$\,yr$^{-1}$, and it lies outside the typical range for T\,Tauri stars of 10$^{-6}$ to 10$^{-8}$\,M$_{\sun}$\,yr$^{-1}$. Nevertheless, our result compares well with a previous determination from Plateau de Bure interferometric data of 1.0~10$^{-9}$\,M$_{\sun}$\,yr$^{-1}$ \citep{Pety06}. Our results are also in agreement with the estimate of 1.7~10$^{-9}$\,M$_{\sun}$\,yr$^{-1}$, made assuming a jet velocity of 200\,km\,s$^{-1}$ \citep{Bacciotti99}. No angular momentum outflow rate could be calculated for this target due to lack of evidence of toroidal velocities \citep{Coffey07}. 

\subsection{CW\,Tau approaching jet}
As in the case of the DG\,Tau jet, HVC and LVC are clearly separated in CW\,Tau jet. The mass outflow rate is found to be almost evenly distributed between the two components giving values of 0.4 and 0.3~10$^{-8}$\,M$_{\sun}$\,yr$^{-1}$ for HVC and LVC respectively but here, unlike DG\,Tau, the datapoints are few and so the result is less reliable. When compared with the mass accretion rate of 10$^{-6}$\,M$_{\sun}$\,yr$^{-1}$ \citep{Hartigan95}, we obtain a ratio of $\sim$0.01. As in all cases where the sulfer doublet ratio saturates, the mass outflow rate of the jet may be underestimated. The angular momentum outflow rate in the approaching jet, also evenly distributed in velocity, was found to be 0.5 and 0.6~10$^{-6}$\,M$_{\sun}$\,yr$^{-1}$\,AU\,km\,s$^{-1}$ for the HVC and LVC respectively. The total of 1.1~10$^{-6}$\,M$_{\sun}$\,yr$^{-1}$\,AU\,km\,s$^{-1}$ represents the lowest angular momentum outflow rate of all our targets. 

\subsection{RW\,Aur receding jet}
Despite the lack of datapoints, 
a mass outflow rate of 1.7~10$^{-8}$\,M$_{\sun}$\,yr$^{-1}$ 
was estimated at 0$\farcs$3 above the disk-plane. 
This value is lower than 3.0~10$^{-8}$\,M$_{\sun}$\,yr$^{-1}$ 
reported at 0$\farcs$2 above the disk-plane from HST/STIS 
spectra with the slit parallel to the flow direction (\citealp{Woitas02}). 
As in the case of DG\,Tau, the difference arises from the fact 
that the latter calculation adopted a jet radius 
according to the slit position rather than the jet FWHM. 
Their jet radius is 1.5 times our measurement. 
Our mass outflow rate result, when doubled to account 
for a bipolar flow, is 2$\%$ of the mass accretion flux 
of 1.6~10$^{-6}$\,M$_{\sun}$\,yr$^{-1}$ \citep{Hartigan95}. 
We estimate the angular momentum outflow rate in the receding jet 
to be 2.9~10$^{-6}$\,M$_{\sun}$\,yr$^{-1}$\,AU\,km\,s$^{-1}$, 
compared to 1.0~10$^{-5}$\,M$_{\sun}$\,yr$^{-1}$\,AU\,km\,s$^{-1}$ 
estimated for the aforementioned parallel slit data \citep{Woitas02}. 
Again, the difference arises from the difference in radii. 
The estimates increase to 6.0~10$^{-8}$\,M$_{\sun}$\,yr$^{-1}$ and 
2.6$^{-5}$\,M$_{\sun}$\,yr$^{-1}$\,AU\,km\,s$^{-1}$ 
for measurements further above the disk-plane 
at 0$\farcs$5 \citep{Woitas05}.   

\begin{table*}
\begin{center}
\scriptsize{\begin{tabular}{lccccc}
\tableline\tableline
Target				&FWHM 	&$\overline{v_p}$&$\dot M_{jet}$&$\overline{v_\phi}$	&$\dot L_{jet}$	\\
				&(arcsec)&($km\,s^{-1}$)	&(10$^{-8}$\,M$_\sun$\,yr$^{-1}$)	&($km\,s^{-1}$)	&(10$^{-6}$\,M$_\sun$\,yr$^{-1}$\,AU\,km\,s$^{-1}$) \\ 
\tableline
TH\,28 receding jet		&0.32	&165		&1.2		&8	&2.5	\\ 
DG\,Tau approaching jet HVC	&0.14	&224		&4.1		&20	&8.4	\\
DG\,Tau approaching jet LVC	&0.20	&97		&2.6		&12	&4.5	\\
HH\,30 approaching jet		&0.20	&54		&0.4		&...	&...	\\
CW\,Tau approaching jet HVC	&0.17	&148		&0.4		&10	&0.5	\\
CW\,Tau approaching jet LVC	&0.35	&42		&0.3		&7	&0.6	\\
RW\,Aur receding jet		&0.18	&129		&1.7		&14	&2.9	\\
\tableline
\end{tabular}}
\end{center}
\caption{Mass, $\dot M_{jet}$, and angular momentum outflow rates, $\dot L_{jet}$, for each jet target. 
Mass flux is integrated over the full jet cross-section. 
Values are divided into higher velocity component (HVC), and lower velocity component (LVC) where applicable. 
\label{massflux}}
\end{table*}
\section{Conclusions}
\label{conclusions}

We have analysed the gas physics for several T\,Tauri jets close to the launching point, based on 
high resolution {\it HST}/STIS spectra taken with the slit perpendicular to the flow direction. 
We have applied the {\it BE diagnostic technique} to the line emission spectra, 
to obtain maps of the electron density, temperature and ionisation fraction. 
To this aim, we have also adapted the {\it BE code}, to allow inclusion of 
more datapoints in the calculations. The new approach involves easing the algebraic constraints, 
which cause difficulties where jets exhibit a broad range of velocities and the various the various lines trace different velocity components, i.e. DG\,Tau. 
Our results represent the first survey of physical conditions at the base of T Tauri jets 
presented in the form of position-velocity diagrams for the physical quantities. 
In fact, with our high spatial and spectral resolution dataset, we resolved the jet physics as a function of 
velocity and distance from the jet axis, in the region just a few tens of AU above the disk plane where the 
flow is launched. 

The overall survey results indicate that, at the jet base, 
the plasma has a high electron density ($>$2~10$^{4}$\,cm$^{-3}$), 
high electron temperature (2\,10$^{4}$\,K), 
and low ionisation level (0.03 - 0.3) 
which varies considerably depending on the target. 
Indeed, in all cases saturation of the [\ion{S}{2}] doublet 
is reached in at least some of the datapoints, 
and so in this region close to the launch point 
of the jet the electron density reported is often a lower limit. 
This is different to findings further along the jet which are 
a factor of ten lower (\citealp{Bacciotti99}; \citealp{Podio06}). 
In the case where previous studies have been carried out, we find good agreement for values 
reported close to the star. 
In the case of TH\,28 (which presents the best dataset), 
possible shock signatures are present, 
thus providing a observational indications that 
shocks can contribute to heating the jet close to the source. 

We determine the mass and angular momentum outflow rates for the jets close 
to their base. Estimates we determined for the mass and angular momentum outflow rates, 
both of which are fundamental parameters in constraining models of accretion/ejection structures, 
particularily if the parameters can be determined close to the jet footpoint. 
Values for a single jet lobe are in the range 
4.0~10$^{-9}$ to 6.7~10$^{-8}$\,M$_\sun$\,yr$^{-1}$ and 
1.1~10$^{-6}$ to 1.3~10$^{-5}$\,M$_\sun$\,yr$^{-1}$\,AU\,km\,s$^{-1}$. 
Again we find good agreement with the 
literature in cases where values were reported close to the star. 
Mass flow ratios were found to be $\dot M_{jet}/\dot M_{acc}$\,$\sim$\,0.01 - 0.07 
where accretion rates were available in the literature (i.e. for 3 of 5 targets). 
This is in the range predicted 
by accretion-ejection models (\citealp{Konigl00}; \citealp{Casse00}; \citealp{Shu00}). 

Although we have examined a region of the jet at about 
50 - 80\,AU from the source corresponding to the collimation zone, 
we note that the region where the jet is formed and launched is believed to 
be on scales of less than 1\,AU, a region which is currently out of the reach of 
present instrumentation and often obscured by infalling matter. 
We await near-infrared interferometry as an opportunity to observing this zone. 

\vspace {0.2in}
{\bf Acknowledgements} 
\vspace {0.1in}
\newline
The present work was supported in part by the European Community's Marie Curie Actions - Human Resource and Mobility within the JETSET (Jet Simulations, Experiments and Theory) network, under contract MRTN-CT-2004-005592. 



\begin{thebibliography}{}

\bibitem[Appenzeller~et~al.(2005)]{Appenzeller05}
Appenzeller, A., Bertout, C., \& Stahl, O., 2005, A\&A, {\bf 434}, 1005

\bibitem[Bacciotti~et~al.(1995)]{Bacciotti95}
Bacciotti F., Chiuderi C., Oliva E., 1995, A\&A, {\bf 296}, 185

\bibitem[Bacciotti~et~al.(1996)]{Bacciotti96}
Bacciotti, F., Hirth, G. A., Natta, A., 1996, A\&A, {\bf 310}, 309

\bibitem[Bacciotti\,\&\,Eisl\"offel(1999)]{BE99} 
Bacciotti, F., \& Eisl\"offel, J., 1999, A\&A, {\bf 342}, 717

\bibitem[Bacciotti~et~al.(1999)]{Bacciotti99} 
Bacciotti, F., Eisl\"offel, J., Ray, T. P., 1999, A\&A, {\bf 350}, 917

\bibitem[Bacciotti~et~al.(2000)]{Bacciotti00}
Bacciotti, F., Mundt, R., Ray, T. P., Eisl\"offel, J., Solf, J., Camenzind, M., 2000, \apj, {\bf 537L}, 49

\bibitem[Bacciotti~et~al.(2002)]{Bacciotti02} 
Bacciotti, F., Ray, T. P., Mundt, R., Eisl\"offel, J., Solf, J., 2002, \apj, {\bf 576}, 222

\bibitem[Bacciotti(2002)]{Bacciotti02mexico}
Bacciotti, F., 2002, RMxAC, 13, 8

\bibitem[Bally~et~al.(2007)]{Bally07}
Bally, J., Reipurth, B., \& Davis, C. J., 2007, in Protostars \& Planets V, B. Reipurth, D. Jewitt \& K. Keil (Tuscon: Univ. Arizona Press), 215

\bibitem[Beckwith~et~al.(1990)]{Beckwith90}
Beckwith, S. V. W., Sargent, A. I., Chini, R. S., Guesten, R., 1990, AJ, {\bf 99}, 924

\bibitem[Burrows~et~al.(1996)]{Burrows96}
Burrows, C. J., Stapelfeldt, K. R., Watson, A. M., Krist, J. E., Ballester, G. E., Clarke, J. T., Crisp, D., Gallagher III, J. S., Griffiths, R. E., Hester, J. J., Hoessel, J. G., Holtzman, J. A., Mould, J. R., Scowen, P. A., Trauger, J. T., \& Westphal, J. A.,  1996, ApJ, {\bf 473}, 437

\bibitem[Casse~\&~Ferriera(2000)]{Casse00}
Casse~\&~Ferriera, 2000, A\&A, 353, 1115

\bibitem[Coffey~et~al.(2004)]{Coffey04}
Coffey, D., Bacciotti, F., Woitas, J., Ray, T. P., \& Eisl\"offel, J., 2004, \apj, {\bf 604}, 758

\bibitem[Coffey~et~al.(2007)]{Coffey07}
Coffey, D., Bacciotti, F., Ray, T. P., Eisl\"offel, J., \& Woitas, J., 2007, \apj, {\bf 663}, 350

\bibitem[Dougados~et~al.(2000)]{Dougados00}
Dougados, C., Cabrit, S., Lavalley-Fouquet, C., M\'enard, F., 2000, A\&A, {\bf 357}, 61

\bibitem[Dougados~et~al.(2002)]{Dougados02}
Dougados, C., Cabrit, S., Lavalley-Fouquet, C., 2002, RevMexAA, {\bf 13}, 43

\bibitem[Eiroa~et~al.(2002)]{Eiroa02} 
Eiroa, C., Oudmaijer, R. D., Davies, J. K., de Winter, D., Garz\'on, F., Palacios, J., Alberdi, A., Ferlet, R., Grady, C. A., Cameron, A., Deeg, H. J., Harris, A. W., Horne, K., Merín, B., Miranda, L. F., Montesinos, B., Mora, A., Penny, A., Quirrenbach, A., Rauer, H., Schneider, J., Solano, E., Tsapras, Y., Wesselius, P. R., 2002, \aap, {\bf 384}, 1038

\bibitem[Eisl\"{o}ffel~et~al.(1998)]{Eisloffel98}
Eisl\"{o}ffel, J., \& Mundt, R., 1998, \apj, {\bf 115}, 1554

\bibitem[Ghez~et~al.(1997)]{Ghez97} 
Ghez, A.~M., White, R.~J., Simon, M., 1997, \apj, {\bf 490}, 353

\bibitem[G\'omez\,de\,Castro(1993)]{GomezdeCastro93}
G\'omez de Castro, A. I., 1993, \apj, {\bf 412}, 43

\bibitem[Graham \& Heyer(1988)]{Graham88}
Graham, J. A. \& Heyer, M. H., 1988, PASP, {\bf 100}, 1529

\bibitem[Hartmann~et~al.(1986)]{Hartmann86}
Hartmann, L., Hewett, R., Stahler, S. \& Mathieu R. D., 1986, \apj, {\bf 309}, 275

\bibitem[Hartigan~et~al.(1987)]{Hartigan87}
Hartigan, P., Raymond, J. \& Hartmann, L., 1987, \apj, {\bf 316}, 323

\bibitem[Hartigan~et~al.(1994)]{Hartigan94}
Hartigan, P., Morse, J. A., Raymond, J., 1994, \apj, {\bf 436}, 125

\bibitem[Hartigan~et~al.(1995)]{Hartigan95}
Hartigan, P., Edwards, S., \& Gandhour, L., 1995, \apj, {\bf 452}, 736

\bibitem[Hartigan~et~al.(2004)]{Hartigan04}
Hartigan, P., Edwards, S. \& Pierson, R., 2004, \apj, {\bf 609}, 261

\bibitem[Hartigan~et~al.(2007)]{Hartigan07}
Hartigan, P., \& Morse, J., 2007, \apj, {\bf 660}, 426

\bibitem[Hern\'{a}ndez~et~al.(2004)]{Hernandez04}
Hern\'{a}ndez, J., Calvet, N., Brice\~{n}o, C., Hartmann, L., Berlind, P., 2004, AJ, {\bf 127}, 1682

\bibitem[Hirth~et~al.(1994)]{Hirth94}
Hirth, F., Mundt, R., Solf, J., Ray, T. P., 1994, ApJ, {\bf 427L}, 99

\bibitem[Hudson~et~al.(2005)]{Hudson05} 
Hudson 2005

\bibitem[Keenan~et~al.(1996)]{Keenan96}
Keenan 1996

\bibitem[K\"{o}nigl~\&~Pudritz(2000)]{Konigl00}
K\"{o}nigl, A., \& Pudritz, R. E., 2000, Protostars and Planets IV, 759

\bibitem[Krautter(1986)]{Krautter86}
Krautter, J., 1986, A$\&$A, {\bf 161}, 195

\bibitem[Lavalley-Fouquet~et~al.(2000)]{Lavalley-Fouquet00}
Lavalley-Fouquet, C., S. Cabrit, S., Dougados, C., A\&A, 2000, {\bf 356}, 41

\bibitem[L\'{o}pez-Mart\'{i}n~et~al.(2003)]{Martin03}
L\'{o}pez-Mart\'{i}n, L., Cabrit, S. \& Dougados, C., 2003, A$\&$A, {\bf 405}, L1

\bibitem[Mendoza~et~al.(1983)]{Mendoza83}
Mendoza 1983

\bibitem[Melnikov~et~al.(2008)]{Melnikov08}
Melnikov, S., Woitas, J., Ray., T. P., Bacciotti, F., Eisl\"{o}ffel, J., 2008, A\&A, 483, 199

\bibitem[Mundt\,\&\,Fried(1983)]{Mundt83}
Mundt, R., \& Fried, R. W., 1983, \apj, {\bf 274L}, 83 

\bibitem[Mundt~et~al.(1990)]{Mundt90}
Mundt, R., Buehrke, T., Solf, J., Ray, T. P. \& Raga, A. C., 1990, A$\&$A, {\bf 232}, 37

\bibitem[Nisini~et~al.(2005)]{Nisini05}
Nisini, B., Bacciotti, F., Giannini, T., Massi, F., Eisl\"{o}ffel, J., Podio, L., Ray, T. P., 2005, A\&A, {\bf 441}, 159

\bibitem[Osterbrock(1989)]{Osterbrock89}
Osterbrock, D. E., 1989, SvA, {\bf 33}, 694

\bibitem[Osterbrock(1994)]{Osterbrock94}
Osterbrock 1994

\bibitem[Pesenti~et~al.(2004)]{Pesenti04}
Pesenti, N., Dougados, S., Cabrit, S., Ferreira, J., Casse, F., Garcia, P., \& O'Brien, D., 2004, A\&A, {\bf 416}, L9

\bibitem[Pety~et~al.(2006)]{Pety06} 
Pety, J., Gueth, F., Guilloteau, S., \& Dutrey, A., 2006, A\&A, {\bf 458}, 841

\bibitem[Podio~et~al.(2006)]{Podio06} 
Podio, L., Bacciotti, F., Nisini, B., Eisl\"{o}ffel, J., Massi, F., Giannini, T., Ray, T. P., 2006, A\&A, {\bf 456}, 189

\bibitem[Pudritz~et~al.(2007)]{Pudritz07}
Pudritz, R. E., Ouyed, R., Fendt, C., \& Brandenburg, A., 2007, in Protostars \& Planets V, B. Reipurth, D. Jewitt \& K. Keil (Tuscon: Univ. Arizona Press), 277

\bibitem[Raga(1992)]{Raga92}
Raga 1992

\bibitem[Ray~et~al.(2007)]{Ray07}
Ray, T. P., Dougados, C., Bacciotti, F., Eisl\"{o}ffel, J., \& Chrysostomou, A., 2007, in Protostars \& Planets V, B. Reipurth, D. Jewitt \& K. Keil (Tuscon: Univ. Arizona Press), 231

\bibitem[Shang~et~al.(2007)]{Shang07}
Shang, H., Li, Z.-Y., \& Hirano, N., 2007, in Protostars \& Planets V, B. Reipurth, D. Jewitt \& K. Keil (Tuscon: Univ. Arizona Press), 261

\bibitem[Shu~et~al.(2000)]{Shu00}
Shu, F. H., Najita, J. R., Shang, H. \& Li, Z.-Y., 2000, in Protostars and Planets IV, V. Mannings, A. P. Boss, $\&$ S. S. Russell (Tuscon: Univ. Arizona Press), 789

\bibitem[Woitas~et~al.(2001)]{Woitas01}
Woitas, J., K{\" o}hler, R., \& Leinert, C., 2001, AAp, {\bf 369}, 249

\bibitem[Woitas~et~al.(2002)]{Woitas02}
Woitas, J,  Ray, T. P., Bacciotti, F, Davis, C, J., Eisl\"offel, J., 2002, \apj, {\bf 580}, 336

\bibitem[Woitas~et~al.(2005)]{Woitas05}
Woitas, J., Bacciotti, F., Ray, T. P., Marconi, A., Coffey, D. \& Eisl\"offel, J., 2004, A$\&$A, {\bf 432}, 149

\end{thebibliography}
\end{document}